\documentclass[aps,prb,eqsecnum,twocolumn]{revtex4}
\ifx\pdftexversion\undefined
  \usepackage[dvips]{graphicx}
  \DeclareGraphicsExtensions{.eps}
\else
  \usepackage[pdftex]{graphicx}
  \DeclareGraphicsExtensions{.pdf}
\fi

\usepackage{float}
\usepackage{amsfonts}
\usepackage{amsmath}
\usepackage{amssymb}

\makeatletter

\providecommand{\tabularnewline}{\\}

\makeatother
\begin{document}

\preprint{This line only printed with preprint option}

\title{Path integral evaluation of equilibrium isotope effects}

\author{Tom\'{a}\v{s} Zimmermann}

\author{Ji\v{r}\'{\i} Van\'{\i}\v{c}ek}

\email{jiri.vanicek@epfl.ch}

\affiliation{Laboratory of Theoretical Physical Chemistry, Institut des Sciences
et Ing\'{e}nierie Chimiques, Ecole Polytechnique F\'{e}d\'{e}rale de Lausanne
(EPFL), CH-1015, Lausanne, Switzerland}

\begin{abstract}
A general and rigorous methodology to compute the quantum equilibrium
isotope effect is described. Unlike standard approaches, ours does
not assume separability of rotational and vibrational motions and
does not make the harmonic approximation for vibrations or rigid rotor
approximation for the rotations. In particular, zero point energy
and anharmonicity effects are described correctly quantum mechanically.
The approach is based on the thermodynamic integration with respect
to the mass of isotopes and on the Feynman path integral representation
of the partition function. An efficient estimator for the derivative
of free energy is used whose statistical error is independent of the
number of imaginary time slices in the path integral, speeding up
calculations by a factor of $\sim60$ at $500$ $\mathrm{K}$ and
more at room temperature. We describe the implementation of the methodology
in the molecular dynamics package Amber 10. The method is tested on
three {[}1,5] sigmatropic hydrogen shift reactions. Because of the
computational expense, we use ab initio potentials to evaluate the
equilibrium isotope effects within the harmonic approximation, and
then the path integral method together with semiempirical potentials
to evaluate the anharmonicity corrections. Our calculations show that
the anharmonicity effects amount up to $30\%$ of the symmetry reduced
reaction free energy. The numerical results are compared with recent
experiments of Doering and coworkers, confirming the accuracy of the
most recent measurement on 2,4,6,7,9-pentamethyl-5-(5,5-$^{2}$H$_{2}$)methylene-11,11a-dihydro-12\textit{H}-naphthacene 
as well as concerns about compromised accuracy, due to side reactions, 
of another measurement on 2-methyl-10-(10,10-$^{2}$H$_{2}$)methylenebicyclo{[}4.4.0]dec-1-ene.
\end{abstract}



\maketitle

\section{Introduction\label{sec:Introduction}}

The equilibrium (thermodynamic) isotope effect (EIE) is defined as
the effect of isotopic substitution on the equilibrium constant. Denoting
an isotopolog with a lighter (heavier) isotope by a subscript \textit{l}
(\textit{h}), the EIE is defined as the ratio of equilibrium constants\begin{equation}
\mathrm{EIE}=\frac{K_{l}}{K_{h}}=\frac{Q_{l}^{\left(p\right)}/Q_{l}^{\left(r\right)}}{Q_{h}^{\left(p\right)}/Q_{h}^{\left(r\right)}},\label{eq:EIE_def}\end{equation}
where $Q^{\left(r\right)}$ \textit{}and $Q^{\left(p\right)}$ are
molecular partition functions per unit volume of reactant and product.
We study a specific case of EIE - the equilibrium ratio of two isotopomers.
In this case, the EIE is equal to the equilibrium constant of the
isotopomerization reaction, \begin{equation}
\mathrm{EIE}=K_{\mathrm{eq}}=\frac{Q^{\left(p\right)}}{Q^{\left(r\right)}},\label{eq:EIE}\end{equation}
where superscripts \textit{r} and \textit{p} refer to reactant and
product isotopomers, respectively. 

Usually, equilibrium isotope effects are computed only approximately:\cite{Alston1996,Anbar2005825,Gawlita1994,Hill20081939,Hrovat1997,Janak2003a,Kolmodin2002,Munoz-Caro2003,Ruszczycky2009107,Saunders2007,Slaughter2000,Smirnov2009,Zeller2002}
In particular, effects due to indistinguishability of particles and
rotational and vibrational contributions to the EIE\ are treated
separately. Furthermore, the vibrational motion is approximated by
a simple harmonic oscillator and the rotational motion is approximated
by a rigid rotor. In general, none of the contributions, not even
the indistinguishability effects can be separated from the others.\cite{Pollock1987,Boninsegni1995}
However, at room temperature or above the nuclei can be accurately
treated as distinguishable, and the indistinguishability effects can
be almost exactly described by symmetry factors. On the other hand,
the effective coupling between rotations and vibrations, anharmonicity
of vibrations, and non-rigidity of rotations can in fact become more
important at higher temperatures. For simplicity, from now on we denote
these three effects together as {}``anharmonicity effects\textquotedblright\ and
the approximation that neglects them the {}``harmonic approximation\textquotedblright\ (HA).
In some cases the effects of anharmonicity of the Born-Oppenheimer
potential surface on the value of EIE can be substantial.\cite{TorresL._jp001327p}
Ishimoto \textit{et al.} have shown that the isotope effect on certain
barrier heights%
\footnote{Namely, the barrier height of internal hindered rotation of the methyl
group.%
} can even have opposite signs, when calculated taking anharmonicity
effects into account and in the HA.\cite{Ishimoto2008} 

Our goal is to describe rigorously equilibria at room temperature
or above. Therefore, two approximations that we make are the Born-Oppenheimer
approximation and the distinguishable particle approximation (we treat
indistinguishability by appropriate symmetry factors). The error due
to the Born-Oppenheimer approximation was studied for H/D EIE by Bardo
and Wolfsberg,\cite{Bardo1976} and by Kleinman and Wolfsberg,\cite{Kleinman1973}
and was shown to be of the order of 1 \% in most studied cases. Since
we assume that nuclei are point charges, the Born-Oppenheimer approximation
implies that the potential energy surface is the same for the two
isotopomers. The differences of Born-Oppenheimer surfaces due to differences
of nuclear volume and quadrupole of isotopes can be important for
heavy elements\cite{Bigeleisen1996,Knyazev19992579,Cowan1981}, but
these are not studied in this work. 

Symmetry factors themselves result in an EIE equal to a rational ratio,
which can be computed analytically. In order to separate the symmetry
contributions from the mass contributions to the EIE, it is useful
to introduce the reduced reaction free energy,\begin{equation}
\Delta F^{\mathrm{red}}=-k_{B}T\ln K_{\mathrm{eq}}^{\mathrm{red}}=-k_{B}T\ln\left(\frac{s^{(r)}Q^{(p)}}{s^{(p)}Q^{(r)}}\right),\label{eq:F_red}\end{equation}
where $s^{(r)}$ and $s^{(p)}$ are the symmetry factors discussed
in more detail in Sec. \ref{sec:Examples}. To include the effects
of quantization of nuclear degrees of freedom beyond the HA rigorously
we use the Feynman path integral representation (PI) of the partition
function. The quantum reduced reaction free energy can then be computed
by thermodynamic integration with respect to the mass of the isotopes.
To compute the derivative of the free energy efficiently, we use a
generalized virial estimator. The advantage of this estimator is that
its statistical error does not increase with the number of imaginary
time slices in the discretized path integral. As a consequence, converged
results can be obtained in a significantly shorter simulation than
with other estimators.

The ultimate goal would be to combine the path integral methodology
with \textit{ab initio} potentials. However, since millions of samples
are required, the computational expense results in the following {}``compromise:\textquotedblright\ First,
the equilibrium isotope effects are computed using \textit{ab initio}
potentials, but as usual, within the HA. Then all anharmonicity corrections
are computed using the PI methodology, but with semiempirical potentials.
In other words, we take advantage of the higher accuracy of the \textit{ab
initio} potentials to compute the harmonic contribution to the EIE
and then make an assumption that the anharmonicity effects are similar
for \textit{ab initio} and semiempirical potentials.

After describing theoretical features of the method, we apply it to
hydrocarbons used in experimental measurements of isotope effects
on {[}1,5] sigmatropic hydrogen transfer reactions. Two of them were
recently used by Doering \textit{et al.} \cite{Doering2006,Doering2007}
who reported equilibrium ratios of their isotopomers. This allows
us to validate our calculations as well as to discuss the apparent
discrepancy in measurements of Doering \textit{et al.} from a theoretical
point of view.

The outline of the paper is as follows: In Sec. \ref{sec:Method},
we describe a rigorous quantum-mechanical methodology to compute the
EIE. Section \ref{sec:Examples} presents the {[}1,5] sigmatropic
hydrogen shift reactions on which we test the methodology, explains
how \textit{ab initio} methods can be combined with the PI\ to compute
the EIE, and discusses in detail symmetry effects in these reactions.
Section \ref{sec:Computational-details} explains the implementation
of the method in Amber 10 and describes details of calculations and
error analysis of our PIMD simulations. Results of calculations are
presented and compared with experiments in Sec. \ref{sec:Results}.
Section \ref{sec:Discussion-and-conclusions} concludes the paper.

\section{The methodology\label{sec:Method}}

\subsection*{Thermodynamic integration}

EIE can be calculated by a procedure of thermodynamic integration\cite{Chandler:1987}
with respect to the mass. This method takes advantage of the relationship\begin{equation}
\mathrm{EIE}=\frac{Q^{(p)}}{Q^{(r)}}=\exp\left[-\beta\int_{0}^{1}d\lambda\frac{dF\left(\lambda\right)}{d\lambda}\right],\label{eq:TI}\end{equation}
where $F=-\log Q/\beta$ is the (quantum) free energy and $\lambda$
is a parameter which provides a smooth transition between isotopomers
$r$ and $p$. This can be accomplished, e.g., by linear interpolation
of masses of all atoms in a molecule according to the equation \begin{equation}
m_{i}\left(\lambda\right)=\left(1-\lambda\right)m_{i}^{(r)}+\lambda m_{i}^{(p)}.\label{eq:lambda_mass}\end{equation}
 In contrast to the partition function itself, the integrand of Eq.
\eqref{eq:TI} \begin{equation}
\frac{dF\left(\lambda\right)}{d\lambda}=-\frac{1}{\beta}\frac{d\log Q(\lambda)}{d\lambda}=-\frac{1}{\beta}\frac{dQ(\lambda)/d\lambda}{Q(\lambda)}\label{eq:log_derivative}\end{equation}
is a thermodynamic average and therefore can be computed by either
Monte Carlo or molecular dynamics simulations.

\subsection*{Path integral approach}

Classically, the EIE is trivial and Eq. \eqref{eq:TI} can be evaluated
analytically. When quantum effects are important, this simplification
is not possible. To describe quantum thermodynamic effects rigorously,
one can use the path integral formulation of quantum mechanics.\cite{Feynman1965}
In the path integral formalism, thermodynamic properties are calculated
exploiting the correspondence between matrix elements of the Boltzmann
operator and the quantum propagator in imaginary time.\cite{Feynman1965,Pollock1987}
In the last decades, path integrals proved to be very useful in many
areas of quantum chemistry, most recently in calculations of heat
capacities,\cite{Predescu2003} rate constants,\cite{yamamoto:044106} kinetic isotope effects,\cite{vanicek:114309} or diffusion
coefficients.\cite{wang:114708}

Let $N$ \ be the number of atoms, $D$ the number of spatial dimensions,
and $P$ the number of imaginary time slices in the discretized PI
($P=1$ gives classical mechanics, $P\rightarrow\infty$ gives quantum
mechanics). Then the PI representation of the partition function $Q$
in the Born-Oppenheimer approximation is \begin{align}
Q & \simeq V^{-1}C\int d\mathbf{r}^{\left(0\right)}\cdots\int d\mathbf{r}^{\left(P-1\right)}\exp\left[-\beta\Phi\left(\left\{ \mathbf{r}^{\left(s\right)}\right\} \right)\right],\label{eq:pi_qr}\\
C & \equiv\left(\frac{P}{2\pi\hbar^{2}\beta}\right)^{NPD/2}\prod_{i=1}^{N}m_{i}^{PD/2}.\label{eq:pi_prefactor}\end{align}
where $\mathbf{r}^{\left(s\right)}\equiv\left(\mathbf{r}_{1}^{\left(s\right)},\mathbf{r}_{2}^{\left(s\right)},\ldots,\mathbf{r}_{N}^{\left(s\right)}\right)$
is the set of Cartesian coordinates associated with the $s$$^{\text{th}}$
time slice, and $\Phi\left(\left\{ \mathbf{r}^{\left(s\right)}\right\} \right)$
is the effective potential \begin{equation}\begin{split}
\Phi\left(\left\{ \mathbf{r}^{\left(s\right)}\right\} \right)=\frac{P}{2\hbar^{2}\beta^{2}}\sum_{i=1}^{N}m_{i}\sum_{s=0}^{P-1}\left(\mathbf{r}_{i}^{\left(s\right)}-\mathbf{r}_{i}^{\left(s+1\right)}\right)^{2}\\+\frac{1}{P}\sum_{s=0}^{P-1}V\left(\mathbf{r}^{\left(s\right)}\right).\end{split}\label{eq:pi_effpotential}\end{equation}
The $P$ particles representing each nucleus in $P$ different imaginary
time slices are called {}``beads.'' Each bead interacts via harmonic
potential with the two beads representing the same nucleus in adjacent
time slices and via potential $V\left(\mathbf{r}^{\left(s\right)}\right)$
attenuated by factor $1/P$ with beads representing other nuclei in
the same imaginary time slice. By straightforward differentiation
of Eq. \eqref{eq:pi_qr} we obtain the so-called thermodynamic estimator
(TE),\cite{vanicek:054108} \begin{equation}\begin{split}
\frac{dF\left(\lambda\right)}{d\lambda}\simeq-\frac{1}{\beta}\sum_{i=1}^{N}\frac{dm_{i}}{d\lambda}\Bigg\langle \frac{DP}{2m_{i}}\\-\frac{P}{2\hbar^{2}\beta}\sum_{s=0}^{P-1}\left(\mathbf{r}_{i}^{\left(s\right)}-\mathbf{r}_{i}^{\left(s+1\right)}\right)^{2}\Bigg\rangle.\end{split}\label{eq:TE_EST}\end{equation}
A problem with expression \eqref{eq:TE_EST} is that its statistical
error grows with the number of time slices. A similar behavior is
a well known property of the thermodynamic estimator for energy,\cite{herman:5150}
where the problem is caused by the kinetic part of energy. 

\begin{figure}
\includegraphics{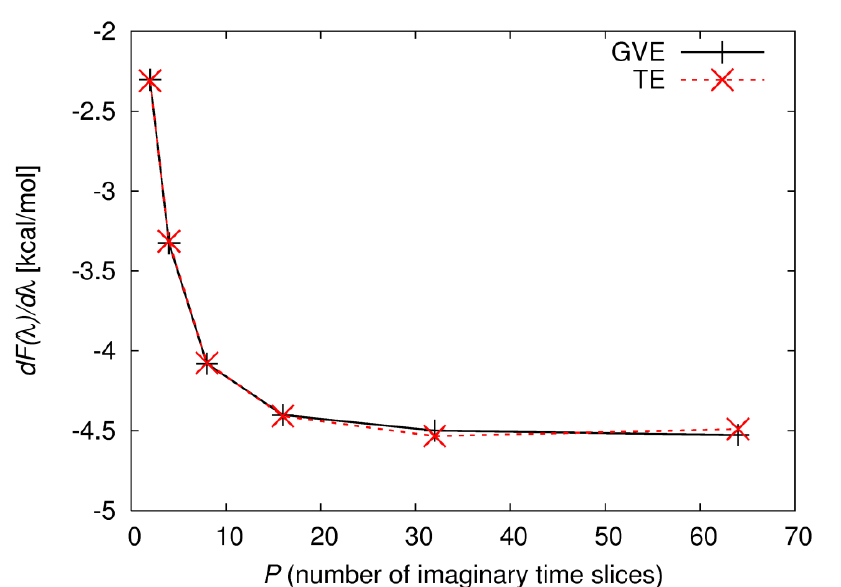}

\caption{Convergence of the generalized virial estimator (GVE) and the thermodynamic
estimator (TE) as a function of the number $P$ of imaginary time
slices in the path integral. All results obtained by 1 $\mathrm{ns}$
long simulations (with the time step 0.05 $\mathrm{fs}$) of compound
\textbf{1-}\textbf{\footnotesize 5,5,5}\textbf{-}\textbf{\textit{d}}\textbf{$_{\text{3}}$}
($\lambda=0$) using GAFF force field, normal mode PIMD, and Nos\'{e}-Hoover
chains of thermostats. \label{fig:GVE_TE_values}}
\end{figure}

\subsection*{Generalized virial estimator }

\begin{figure}
\includegraphics{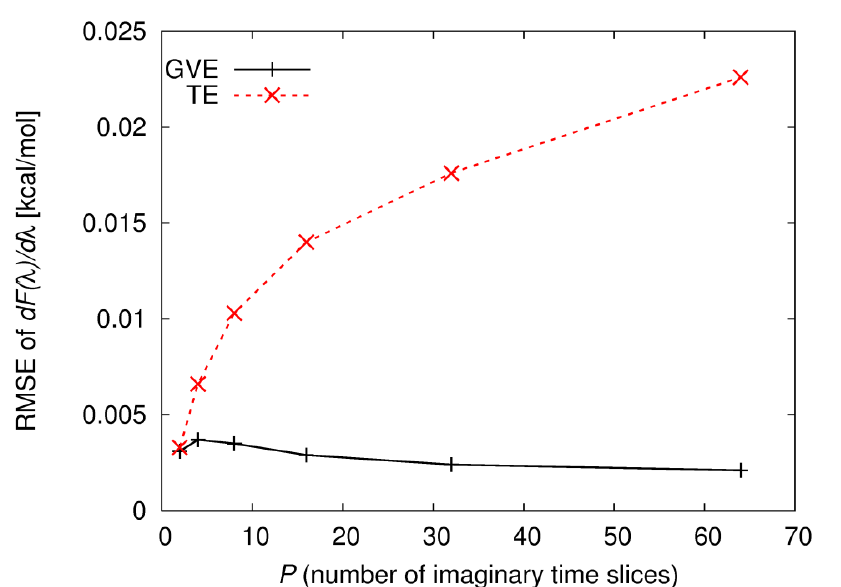}

\caption{Root mean square errors (RMSEs) of the generalized virial estimator
(GVE) and the thermodynamic estimator (TE) as a function of the number
$P$ of imaginary time slices in the path integral. All results obtained
by 1 $\mathrm{ns}$ long simulations (with the time step 0.05 $\mathrm{fs}$)
of compound \textbf{1-}\textbf{\footnotesize 5,5,5}\textbf{-}\textbf{\textit{d}}\textbf{$_{\text{3}}$}
($\lambda=0$) using GAFF force field, normal mode PIMD, and Nos\'{e}-Hoover
chains of thermostats. Note that the RMSE of the GVE is not only non-increasing
(as expected from theory), but in fact decreases slightly with increasing
$P$, which is due to the decrease of correlation length.\label{fig:GVE_TE_errors}}
\end{figure}

The growth of statistical error of the thermodynamic estimator for
energy is removed by expressing the estimator only in terms of the
potential and its derivatives using the virial theorem.\cite{herman:5150}
In our case, a similar improvement can be accomplished, if a coordinate
transformation, \begin{equation}
\mathbf{x}_{i}^{\left(s\right)}=m_{i}^{1/2}(\mathbf{r}_{i}^{\left(s\right)}-\mathbf{r}_{i}^{(C)}),\label{eq:mass_centroid_transf}\end{equation}
is introduced into Eq. \eqref{eq:pi_qr} prior to performing the derivative.
Here, the {}``centroid'' coordinate is defined as

\begin{equation}
\mathbf{r}_{i}^{(C)}=\frac{1}{P}\sum_{s=0}^{P-1}\mathbf{r}_{i}^{(s)}.\label{eq:centroid_coord}\end{equation}
In other words, first the {}``centroid'' coordinate $\mathbf{r}_{i}^{\left(C\right)}$
is subtracted, and then the coordinates are mass-scaled. Resulting
generalized virial estimator (GVE)\cite{vanicek:114309,Vanicek_Dubna_2005}
takes the form

\begin{equation}
\begin{split}
\frac{dF\left(\lambda\right)}{d\lambda}\simeq-\frac{1}{\beta}\sum_{i=1}^{N}\frac{dm_{i}/d\lambda}{m_{i}}\Bigg[\frac{D}{2}\\+\frac{\beta}{2P}\left\langle \sum_{s=0}^{P-1}\left(\mathbf{r}_{i}^{(s)}-\mathbf{r}_{i}^{(C)}\right)\cdot\frac{\partial V\left(\mathbf{r}^{(s)}\right)}{\partial\mathbf{r}_{i}^{(s)}}\right\rangle \Bigg].
\end{split}
\label{eq:gve}
\end{equation}
Its primary advantage is that the root mean square error (RMSE) of
the average, $\sigma_{\mathrm{\text{av}}}$, is approximately independent
on the number of imaginary time slices $P$,\begin{equation}
\sigma_{\mathrm{\text{av}}}\approx O\left(P^{0}\tau_{\mathrm{c}}^{1/2}\tau_{\mathrm{sim}}^{-1/2}\right).\label{eq:sigma_virial}\end{equation}
In this equation $\tau_{\mathrm{sim}}$ denotes the length of the
simulation and $\tau_{\mathrm{c}}$ is the correlation length.

The convergence of values and statistical errors of both estimators
as a function of number $P$ of imaginary time slices for systems
studied in this paper is discussed in Section \ref{sec:Computational-details}.
As expected, up to the statistical error they give the same values
as can be seen in Fig. \ref{fig:GVE_TE_values}. Nevertheless, when
quantum effects are important, and a high value of $P$ must be used,
the GVE is the preferred estimator since it has much smaller statistical
error and therefore converges much faster than the TE (see Fig. \ref{fig:GVE_TE_errors}).

\subsection*{Path integral molecular dynamics}

Thermodynamic average in Eq. \eqref{eq:gve} can be evaluated efficiently
using the path integral Monte Carlo (PIMC) or path integral molecular
dynamics (PIMD). In PIMC, gradients of $V$ in Eq. \eqref{eq:gve}
result in additional calculations since the usual Metropolis Monte
Carlo procedure for the random walk only requires the values of $V$.
This additional cost can be, however, reduced either by less frequent
sampling, or by using a trick in which the total derivative with respect
to $\lambda$ (not the gradients!) is computed by finite difference.\cite{vanicek:114309,Vanicek_Dubna_2005,Predescu2003,Predescu2004}
In case of PIMD, the presence of gradients of $V$ in Eq. \eqref{eq:gve}
does not slow down the calculation since forces are already computed
by a propagation algorithm. Although in principle, a PIMC algorithm
for a specific problem can always be at least as efficient as a PIMD
algorithm, in practice it is much easier to write a general PIMD algorithm
and so PIMD is usually the algorithm used in general software packages.
Since PIMD was implemented in Amber 9,\cite{amber9} we have implemented
the methodology described above for computing EIEs into Amber 10.\cite{amber10}
This implementation is what was used in calculations in the following
sections. In PIMD, equation \eqref{eq:pi_qr} is augmented by fictitious
classical momenta $\boldsymbol{\mathrm{p}}^{\left(s\right)}$,\begin{widetext} \begin{equation}
Q\simeq V^{-1}\tilde{C}\int d\mathbf{p}^{\left(0\right)}\cdots\int d\mathbf{p}^{\left(P-1\right)}\int d\mathbf{r}^{\left(0\right)}\cdots\int d\mathbf{r}^{\left(P-1\right)}\exp\left[-\beta\left({\displaystyle \sum_{s=0}^{P-1}}\frac{\left(\mathbf{p}^{\left(s\right)}\right)^{2}}{2m_{s}}+\Phi\left(\left\{ \mathbf{r}^{\left(s\right)}\right\} \right)\right)\right].\label{eq:pi_qp}\end{equation} \end{widetext}
The normalization prefactor $\tilde{C}$ is chosen in such a way that
the original prefactor $C$ in Eq. \eqref{eq:pi_qr} is reproduced
when the momentum integrals in Eq. \eqref{eq:pi_qp} are evaluated
analytically. The partition function \eqref{eq:pi_qp} is formally
equivalent to the partition function of a classical system of cyclic
polyatomic molecules with harmonic bonds. Each such molecule represents
an individual atom in the original molecule and interacts with molecules
representing other atoms via a potential derived from the potential
of the original molecule.\cite{Chandler1981} This system can be studied
directly using well developed methods of the classical molecular dynamics.

\subsection*{Low- and high-temperature limits}

It is useful to see how the general path integral expressions \eqref{eq:TI},
\eqref{eq:pi_qr}, and \eqref{eq:gve} behave in the low and high
temperature limits. As temperature decreases, the difference of the
reduced free energies of isotopomers approaches the difference of
their zero point energies (ZPEs). Therefore, still assuming that indistinguishability
is correctly described by symmetry factors, the low temperature limit
of the EIE is equal to \begin{equation}
\mathrm{EIE_{low\mathit{T}}}=\frac{s^{\left(r\right)}}{s^{\left(p\right)}}\exp\left[-\beta\left(\varepsilon^{\left(r\right)}-\varepsilon^{\left(p\right)}\right)\right],\label{eq:EIE_low_temp}\end{equation}
where $\varepsilon$ denotes the ZPE.

At high temperature, the system approaches its classical limit. In
this limit, we can set the number of imaginary time slices $P=1$,
and the EIE can be computed analytically using partition function
\eqref{eq:pi_qr}, which becomes\-\-\begin{equation}\begin{split}
Q_{\mathrm{class}}\simeq {\displaystyle V^{-1}\left(\frac{1}{2\pi\hbar^{2}\beta}\right)^{ND/2}\prod_{i=1}^{N}m_{i}^{D/2}}\\\times\int d^{N}\mathbf{r}\exp\left[-\beta V\left(\mathbf{r}_{1},\ldots,\mathbf{r}_{N}\right)\right].\end{split}\label{eq:q_class}\end{equation}
Again, for isotopomers the mass dependent factors cancel out upon
substitution into Eq. \eqref{eq:EIE}, along with all other terms
except for the integrals. Although the Born-Oppenheimer potential
surface is the same for all isotopomers, the integrals do not cancel
out since the integration is not performed over the whole configuration
space, but only over parts attributed to reactants or products. Particles
are treated as distinguishable in the classical limit and the volumes
of the configuration space corresponding to reactants and products
are generally different. After the reduction, EIE becomes exactly
equal to the ratio of the symmetry factors,

\begin{equation}
\mathrm{EIE_{\mathrm{high}\mathit{T}}}=\frac{s^{\left(r\right)}}{s^{\left(p\right)}}.\label{eq:EIE_class}\end{equation}

\section{Equilibrium isotope effects in three {[}1,5] sigmatropic hydrogen
shift reactions\label{sec:Examples}}

\begin{figure}
\includegraphics[keepaspectratio]{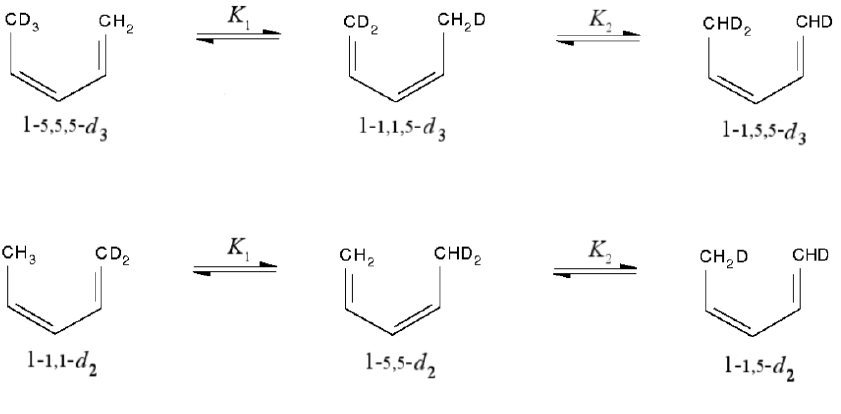}

\caption{Equilibrium of the {[}1,5] hydrogen shift reaction in (3\textit{Z})-(5,5,5-$^{\text{2}}$H$_{\text{3}}$)penta-1,3-diene
(\textbf{1-}\textbf{\footnotesize 5,5,5}\textbf{-}\textbf{\textit{d}}\textbf{$_{\text{3}}$})
and in (3\textit{Z})-(1,1-$^{\text{2}}$H$_{2}$)penta-1,3-diene (\textbf{1-}\textbf{\footnotesize 1,1}\textbf{-}\textbf{\textit{d}}\textbf{$_{\text{2}}$}).
If all contributions except for those due to symmetry factors $s_{a}$
and $s_{b}$ were neglected, one would obtain approximate equilibrium
constants $K_{1}=3$ and $K_{2}=2$ in both cases.\label{fig:equlibrium_pent}}
\end{figure}

We examine the EIE for four related compounds. The parent compound
(3\textit{Z})-penta-1,3-diene (compound \textbf{1}) is the simplest
molecule to model the {[}1,5] sigmatropic hydrogen shift reaction.
Two of its isotopologs, tri-deuterated (3\textit{Z})-(5,5,5-$^{\text{2}}$H$_{\text{3}}$)penta-1,3-diene
(\textbf{1-}\textbf{\footnotesize 5,5,5}\textbf{-}\textbf{\textit{d}}\textbf{$_{\text{3}}$})
and di-deuterated (3\textit{Z})-(1,1-$^{\text{2}}$H$_{2}$)penta-1,3-diene
(\textbf{1-}\textbf{\footnotesize 1,1}\textbf{-}\textbf{\textit{d}}\textbf{$_{\text{2}}$})
(see Fig. \ref{fig:equlibrium_pent}) were used by Roth and K\"onig
to measure an unusually high value of the kinetic isotope effect (KIE)
on the {[}1,5] hydrogen shift reaction with respect to the substitution
of hydrogen by deuterium.\cite{Roth1966} This result pointed to a
significant role of tunneling in {[}1,5] sigmatropic hydrogen transfer
reactions. Subsequently, much theoretical research was devoted to
the study of this reaction. Here we calculate final equilibrium ratios
of products of this reaction, which, to our knowledge, were not theoretically
predicted so far.

\begin{figure*}
\includegraphics{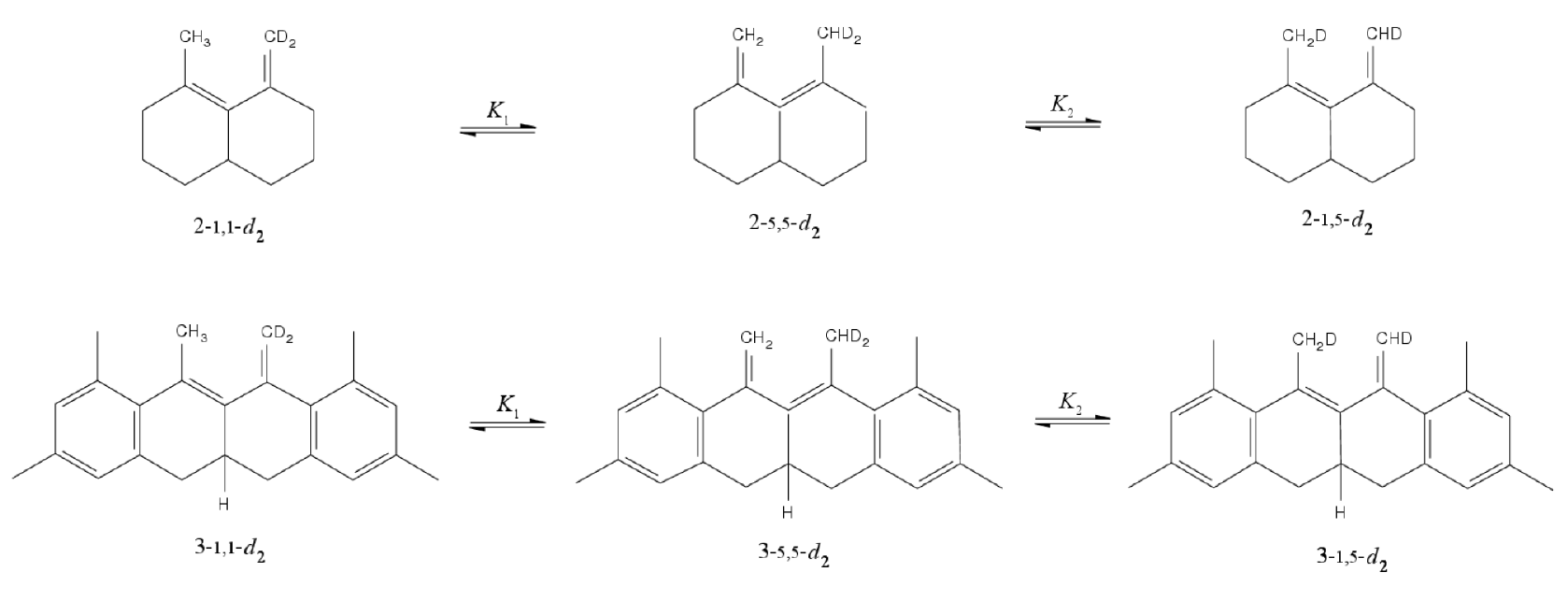}

\caption{Equilibrium of the {[}1,5] hydrogen shift reaction in 2-methyl-10-(10,10-$^{\text{2}}$H$_{2}$)methylenebicyclo{[}4.4.0]dec-1-ene
(\textbf{2-}\textbf{\footnotesize 1,1}\textbf{-}\textbf{\textit{d}}\textbf{$_{\text{2}}$})
and in 2,4,6,7,9-pentamethyl-5-(5,5-$^{\text{2}}$H$_{2}$)methylene-11,11a-dihydro-12\textit{H}-naphthacene
(\textbf{3-}\textbf{\footnotesize 1,1}\textbf{-}\textbf{\textit{d}}\textbf{$_{\text{2}}$})
(the locators of positions of deuterium atoms in abbreviations are
chosen to correspond to locators in (3\textit{Z})-penta-1,3-diene).
As for compound \textbf{1}, values of equilibrium constants imposed
by symmetry are  $K_{1}=3$ and $K_{2}=2$.\label{fig:equilibria_cyclic}}
\end{figure*}

Two other compounds, 2-methyl-10-(10,10-$^{\text{2}}$H$_{2}$)methylenebicyclo{[}4.4.0]dec-1-ene
(\textbf{2-}\textbf{\footnotesize 1,1}\textbf{-}\textbf{\textit{d}}\textbf{$_{\text{2}}$})
and 2,4,6,7,9-pentamethyl-5-(5,5-$^{\text{2}}$H$_{2}$)methylene-11,11a-dihydro-12\textit{H}-naphthacene
(\textbf{3-}\textbf{\footnotesize 1,1}\textbf{-}\textbf{\textit{d}}\textbf{$_{\text{2}}$})
(see Fig. \ref{fig:equilibria_cyclic}), were recently used by Doering
\textit{et al.} to confirm and possibly refine the experimental value
of the KIE on the {[}1,5] hydrogen shift.\cite{Doering2006,Doering2007}
In contrast to (3\textit{Z})-penta-1,3-diene \textbf{(}compound \textbf{1}),
where the s-\textit{trans} conformer incompetent of the {[}1,5] hydrogen
shift is the most stable, pentadiene moiety in compounds 2 and 3 is
locked in the s-\textit{cis} conformation. This not only increases
the reaction rate, but also rules out the (very small) effect of the
EIE on the KIE due to the shift in s-\textit{cis}/s-\textit{trans}
equilibrium. For both molecules Doering \textit{et al.} reported final
equilibrium ratios of isotopomers. Despite the similarity of compounds
\textbf{2} and \textbf{3}, these ratios are qualitatively different.
Indeed, one motivation for measuring EIE in compound \textbf{3} was
that Doering \textit{et al.} suspected that in the case of \textbf{2-}\textbf{\footnotesize 1,1}\textbf{-}\textbf{\textit{d}}\textbf{$_{\text{2}}$}
the equilibrium ratio might be modified by unwanted side reactions,
mainly dimerizations. One of our goals is to elucidate this discrepancy
from the theoretical point of view.

\begin{figure}[b]
\includegraphics{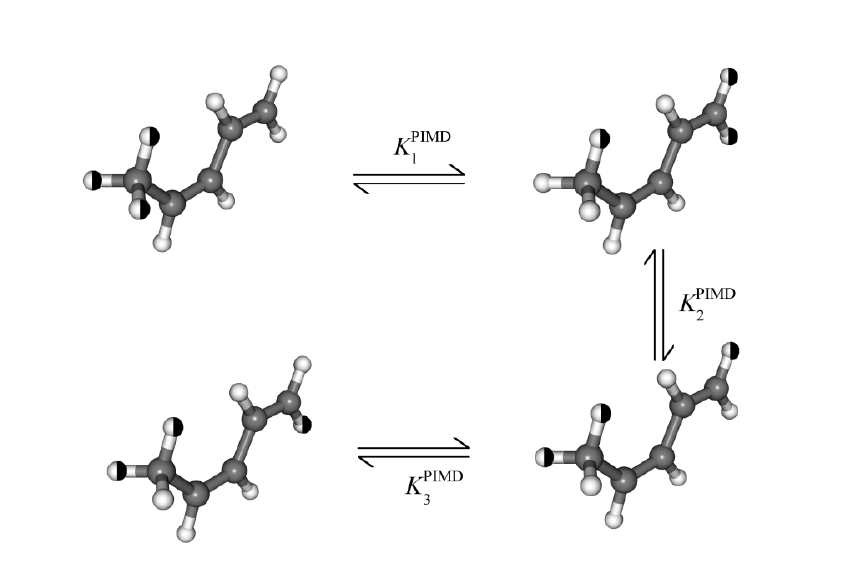}

\caption{Three PIMD simulations used to compute equilibrium ratios of the
{[}1,5] hydrogen shift reaction in (3\textit{Z})-(5,5,5-$^{\text{2}}$H$_{\text{3}}$)penta-1,3-diene
(\textbf{1-}\textbf{\footnotesize 5,5,5}\textbf{-}\textbf{\textit{d}}\textbf{$_{\text{3}}$}).
Half white, half black spheres represent deuterium atoms. The methyl
group, in contrast to methylene, rotates during simulations. \label{fig:PIMD-3steps-pent}}
\end{figure}

As can be seen in Figs. \ref{fig:equlibrium_pent} and \ref{fig:equilibria_cyclic},
the final equilibrium of the {[}1,5] hydrogen shift reaction of all
examined compounds can be described as an outcome of two reactions.
The second reaction (leading from deuterio-methyl-dideuterio-methylene
to dideuterio-methyl-deuterio-methylene in tri-deuterated compounds
and from dideuterio-methyl-methylene to deuterio-methyl-deuterio-methylene
in di-deuterated compounds) produces a mixture of species that differ
by the orientation of the mono-deuterated methylene group. Due to
a high barrier for rotation of this group, the two products cannot
be properly sampled in a single PIMD simulation. Therefore an additional
PIMD simulation is necessary, as shown on the example of \textbf{1-}\textbf{\footnotesize 5,5,5}\textbf{-}\textbf{\textit{d}}\textbf{$_{\text{3}}$}
in Fig. \ref{fig:PIMD-3steps-pent}. The reduced free energy of the
second step is then calculated as \begin{equation}
\Delta F^{\mathrm{red}}=-k_{B}T\ln\left[\frac{K_{2}^{\mathrm{PIMD}}}{2}\left(1+K_{3}^{\mathrm{PIMD}}\right)\right],\label{eq:F_red_2step_pimd}\end{equation}
where 1/2 is the ratio of symmetry factors and $K_{2}^{\mathrm{PIMD}}$
and $K_{3}^{\mathrm{PIMD}}$ stand for equilibrium constants obtained
by the second and third PIMD simulations, respectively. Together they
represent the second reaction step. %

\subsection*{Combination of \textit{ab initio} methods with the PIMD}

Unfortunately, at present the PIMD method cannot be used in conjunction
with higher level \textit{ab initio} methods, due to a high number
of potential energy evaluations needed. Semiempirical methods, which
can be used instead, do not achieve comparable accuracy. We therefore
make the following two assumptions: First, we assume that the main
contribution to the EIE can be calculated in the framework of HA.
Second, we assume that selected semiempirical methods are accurate
enough to reliably estimate the anharmonicity correction. The anharmonicity
correction is calculated as \begin{equation}
\Delta\Delta F^{\mathrm{anharm}}=\Delta F_{\mathrm{PIMD}}^{\mathrm{red}}-\Delta F_{\mathrm{HA}}^{\mathrm{red}}.\label{eq:F_anharm}\end{equation}
 With these two assumptions, we can take advantage of both PIMD and
higher level methods by adding the semiempirical anharmonicity correction
to the HA result calculated by a more accurate method. The HA value
of $\Delta F^{\mathrm{red}}$ is obtained by Boltzmann averaging over
all possible distinguishable conformations,  \begin{equation}
{\displaystyle \Delta F_{\mathrm{HA}}^{\mathrm{red}}=-k_{B}T\ln\left[\frac{s^{(r)}\sum_{i=1}^{N_{p}}\left\{ \exp\left(\frac{-E_{i}^{\mathrm{el}}}{k_{B}T}\right)\sum_{j=1}^{s^{(p)}}Q_{ij}^{\mathrm{\mathit{p,\mathrm{nuc}}}}\right\} }{s^{(p)}\sum_{i=1}^{N_{r}}\left\{ \exp\left(\frac{-E_{i}^{\mathrm{el}}}{k_{B}T}\right)\sum_{j=1}^{s^{(r)}}Q_{ij}^{\mathrm{\mathit{r,\mathrm{nuc}}}}\right\} }\right]},\label{eq:F_red_HA}\end{equation}
 where $N_{r}$ is the number of {}``geometrically different isomers''
of a reactant. By geometrically different isomers we mean species
differing in their geometry, not species differing only in positions
of isotopically substituted atoms. $E_{i}^{\mathrm{el}}$ is the electronic
energy (including nuclear repulsion) of $i^{\mathrm{th}}$ isomer,
$s^{(r)}$ is the symmetry factor and $Q_{ij}^{r,\mathrm{nuc}}$ are
partition functions of the nuclear motion of $s^{(r)}$ isotopomers.
$N_{p}$, $s^{(p)}$, $Q_{ij}^{p,\mathrm{nuc}}$ denote analogous
quantities for the product.

\subsection*{Symmetry factors}

Although symmetry effects can be computed analytically, for the reactions
studied in this paper they are nontrivial and so we discuss them here
in more detail. As mentioned above, we are interested in moderate
temperatures (above $\sim100$ $\mathrm{K}$) where quantum effects
might be very important but the distinguishable particle approximation
remains valid. In this case, effects of particle indistinguishability
and of non-distinguishing several, in principle distinguishable minima,
by an experiment can be conveniently unified by the concept of symmetry
factor. In our setting, we will call {}``symmetry factor'' the product\begin{equation}
s=s^{\mathrm{exp}}\cdot\prod_{i=1}^{N}\frac{1}{\sigma_{i}}.\label{eq:symmetry_factors}\end{equation}
Here, $s^{\mathrm{exp}}$ refers to the number of distinguishable
minima not distinguished by the experiment and $\sigma_{i}$ are the
usual rotational symmetry numbers \textbf{}of symmetric rotors. The
symmetry numbers are present only if either the whole molecule or
some of its parts are treated as free or hindered classical symmetrical
rotors. (In this case, the number of minima of the hindered rotor
potential is not included in $s^{\mathrm{exp}}$.) The symmetry numbers
are not present if rotational barriers are so high that the corresponding
degrees of freedom should be considered as vibrations. 

The concept can be illustrated on an example of the mono- and non-deuterated
methyl group in a rotational potential with three equivalent minima
120$\,^{\circ}$ apart. At low temperatures, when hindered rotation
of the methyl group reduces to a vibration, the symmetry factor is
determined only by $s^{\mathrm{exp}}$. In the case of mono-deuterated
methyl group there are three in principle distinguishable minima corresponding
to three rotamers, which are, as we suppose, considered to be one
species by the observer. Therefore $s^{\mathrm{exp}}=3$. In the case
of non-deuterated methyl there is only one distinguishable minimum
and $s^{\mathrm{exp}}=1$. At higher temperatures, when the methyl
group can be treated as a hindered rotor, its contribution to the
symmetry factor is determined only by rotational symmetry number $\sigma$,
where $\sigma=1$ for a mono-deuterated methyl group and $\sigma=3$
for a non-deuterated methyl group. From definition \eqref{eq:symmetry_factors},
it is clear that the high and low temperature pictures are consistent
and give the same ratios of symmetry factors.

Partition functions $Q^{(r)}$ and $Q^{(p)}$ needed in calculation
of the reduced free energy in Eq. \eqref{eq:F_red} are computed as
sums of partition functions of $s^{\mathrm{(r),exp}}$ and $s^{\mathrm{(p),exp}}$
isotopomers.

\section{Computational details\label{sec:Computational-details}}

\subsection*{Ab initio, density functional, and semiempirical methods}

In this subsection, we will discuss the accuracy of four electronic
structure methods used in our calculations. Ab initio MP2 and the
B98 density functional method,\cite{becke:8554,schmider:9624} both
in combination with the 6-311+(2df,p) basis set, were used for calculations
within the HA. Semiempirical AM1\cite{Dewar1985} and SCC-DFTB\cite{Elstner1998}
methods were used in both HA and PIMD calculations. This allowed us
to compute the error introduced by the HA. 

Aside from symmetry factors, the EIE is dominated by vibrational contributions.
Therefore we concentrate mainly on the accuracy of HA vibrational
frequencies. According to Merrick \textit{et al.,}\cite{Merrick2007}
who tested the performance of MP2 and B98 methods by means of comparison
with experimental data for a set of 39 molecules, RMSE of ZPEs is
0.46 $\mathrm{kJ\cdot mol^{-1}}$ at MP2/6-311+(2df,p) and 0.31 $\mathrm{kJ\cdot mol^{-1}}$
at B98/6-311+(2df,p) level of theory. Appropriate ZPE scaling factors
are equal to 0.9777 and 0.9886 respectively. Corresponding RMSEs of
frequencies are 40 $\mathrm{cm^{-1}}$ and 31 $\mathrm{cm^{-1}}.$
Therefore, a slightly higher accuracy can be expected from the B98
functional. The accuracy of both semiempirical methods is significantly
worse than the accuracy of higher level \textit{ab initio} methods.
According to Witek and Morokuma, the RMSE of AM1 frequencies in comparison
to experimental values for 66 molecules is 95 $\mathrm{cm^{-1}}$
with frequency scaling factor equal to 0.9566.\cite{Witek2004} The
error of vibrational frequencies obtained using SCC-DFTB depends on
parametrization. We tested two parameter sets: the original SCC-DFTB
parametrization\cite{Elstner1998} and the parameter set optimized
with respect to frequencies by Malolepsza \textit{et al.}\cite{Malolepsza2005}
(further designated SCC-DFTB-MWM). The error of vibrational frequencies
calculated with the original SCC-DFTB parameters was studied by Kr\"uger
\textit{et al.}\cite{kruger:114110} The mean absolute deviation from
the reference values calculated at BLYP/cc-pVTZ for a set of 22 molecules
was 75 $\mathrm{cm^{-1}}$. The error of the reference method itself,
as compared to an experiment with a slightly smaller set of molecules
was 31 $\mathrm{cm^{-1}}$. In the above mentioned study performed
by Witek and Morokuma, RMSE of 82 $\mathrm{cm^{-1}}$ with scaling
factor 0.9933 was obtained.\cite{Witek2004} The SCC-DFTB-MWM mean
absolute deviation of experimental and calculated frequencies for
a set of 14 hydrocarbons is indeed better and is equal to 33 $\mathrm{cm^{-1}}$
instead of 59 $\mathrm{cm^{-1}}$ for the original parametrization.
\cite{Malolepsza2005} 

The suitability of AM1, SCC-DFTB, SCC-DFTB-MWM, and several other
semiempirical methods for our systems was tested by comparison of
EIE values in the HA and of potential energy scans with the corresponding
quantities computed with MP2 and B98. Results of this comparison are
presented in the appendix. The AM1 method is shown to reproduce \textit{ab
initio} EIEs in the HA very well, but fails to reproduce scans of
potential energies of methyl and vinyl group rotations. Therefore,
in addition to AM1 method we used the SCC-DFTB method, which is, among
the semiempirical methods tested by us, the best in reproducing \textit{ab
initio} potential energy scans. On the other hand, compared to AM1,
SCC-DFTB gives a worse EIE in the HA.

\subsection*{Statistical errors, convergence, and parameters of PIMD simulations}

Statistical RMSEs of averages of PIMD simulations were calculated
from the equation \begin{equation}
\sigma_{\mathrm{av}}=\tau_{\mathrm{c}}^{1/2}\tau_{\mathrm{sim}}^{-1/2}\sigma,\label{eq:correlated_sigma}\end{equation}
where $\sigma$ is the RMSE of one sample. Correlation lengths $\tau_{\mathrm{c}}$
were estimated by the method of block averages.\cite{flyvbjerg:461}
 At constant temperature, the correlation length decreases with increasing
number of imaginary time slices. Also, for constant number of imaginary
time slices, the correlation length decreases with increasing temperature.
Finally, the correlation length stays approximately constant at different
temperatures, if the number of imaginary time slices is chosen so
that $\Delta F$ is converged to approximately the same precision.
In our systems, correlation length is close to 3.5 $\mathrm{ps}$.

The EIE was studied at four different temperatures: 200, 441.05, 478.45,
and 1000 $\mathrm{K}$ using the normal mode version of PIMD.\cite{Berne1986}
To control temperature, the Nos\'{e}-Hoover chains with four thermostats
coupled to each PI degree of freedom were used.\cite{martyna:2635}. 

Different numbers $P$ of imaginary time slices had to be used at
different temperatures, since at lower temperatures quantum effects
become more important, and the number of imaginary time slices necessary
to maintain the desired accuracy increases. To examine the required
value of $P$ as a function of temperature we used the GAFF force
field.\cite{Wang2004} Whereas the accuracy of vibrational frequencies
calculated using the GAFF force field is relatively low {[}RMS difference
of GAFF and B98/6-311+(2df,p) frequencies of compound \textbf{1} is
equal to 125 $\mathrm{cm^{-1}}$], the potential should be realistic
enough for the assessment of the convergence with respect to $P$.
For example, the difference of potential energies $\Delta E$ between
s-\textit{cis} and s-\textit{trans} conformations of compound \textbf{1}
is 4.3 $\mathrm{kcal\cdot mol^{-1}}$ as compared to 2.7 $\mathrm{kcal\cdot mol^{-1}}$
obtained by MP2/6-311+(2df,p) or 3.5 $\mathrm{kcal\cdot mol^{-1}}$
obtained by B98/6-311+(2df,p). 

To check the convergence at 478.45 $\mathrm{K}$, we calculated values
of the integral in Eq. \eqref{eq:TI} for the deuterium transfer reaction
in \textbf{1-}\textbf{\footnotesize 5,5,5}\textbf{-}\textbf{\textit{d}}\textbf{$_{\text{3}}$}
with 40 and 48 imaginary time slices (using Simpson's rule with 5
points). Their difference is equal to 0.00005 $\pm$ 0.00040 $\mathrm{kcal\cdot mol^{-1}}$.
This is less than the statistical error of the calculation on the
model system, which itself is smaller than the error of production
calculations, since the model calculation was 10 times longer than
the longest production calculation. Therefore, taking into account
the accuracy of production calculations, $P=40$ can be considered
the converged number of imaginary time slices. This is further supported
by the observation of the convergence of the single value of the GVE
\eqref{eq:gve} at $\lambda=0$. The relative difference of GVE values
obtained with 40 and 48 imaginary time slices is equal to 0.22 $\pm$
0.02~\%, whereas the difference between 40 and 72 imaginary time
slices equals to 0.39 $\pm$ 0.04~\%. The discretization error is
asymptotically proportional to $P^{-2}$. By fitting this dependence
to the calculated values, we  estimated the difference between the
values for $P=40$ and for the limit  $P\rightarrow\infty$ to be
less than 0.6~\%. This is only three times more than the difference
between $P=40$ and $P=48$. Therefore, if the integral converges
similarly as the GVE, we can use the aforementioned difference between
40 and 48 imaginary time slices as the criterion of convergence. The
convergence of the GVE with the number of imaginary time slices is
displayed in Fig. \ref{fig:GVE_TE_values}. Since 441.05 $\mathrm{K}$
is close enough to 478.45 $\mathrm{K}$ we used the same value of
$P$ at this temperature. To check the convergence at 200 and 1000
$\mathrm{K}$, we  observed only the convergence of the single value
of the GVE. At 200 $\mathrm{K}$ the relative difference of GVE values
at $\lambda=0$ obtained with $P=72$ and $P=80$ is 0.03 $\pm$ 0.07~\%.
Based on the comparison with the previous result, $P=72$ is considered
sufficient. At 1000 $\mathrm{K}$, the relative difference of GVE
values at $\lambda=0$ for $P=24$ and $P=32$ equals to 0.1 $\pm$
0.02 \%, so that 24 imaginary time slices are used further.

The time step at 441.05 and 478.45 $\mathrm{K}$ was 0.05 $\mathrm{fs}$
to satisfy the requirement of energy conservation. At 200 and 1000
$\mathrm{K}$, a shorter step of 0.025 $\mathrm{fs}$ was used due
to the increased stiffness of the harmonic bonds between beads at
200 $\mathrm{K}$ and due to the increased average kinetic energy
at 1000 $\mathrm{K}$. The simulation lengths differed for different
molecules. For both isotopologs of compound \textbf{1} simulation
length of 1 $\mathrm{ns}$ ensured that the system properly explored
both the s-\textit{trans} and the s-\textit{cis} conformations. Convergence
was checked by monitoring running averages and by comparing the ratio
of the s-\textit{trans} and s-\textit{cis} conformers with the ratio
calculated in the HA. The length of converged PIMD simulations of
compound \textbf{2} was 500 $\mathrm{ps}$. Convergence was checked
again using running averages and by visual analysis of trajectories
to ensure that the system properly explored all local minima. For
compound \textbf{3,} the simulation length was 400 $\mathrm{ps}$. 

The integral in Eq. \eqref{eq:TI} was calculated using the Simpson's
rule. Using the AM1 potential, the GVE was evaluated for five values
of  $\lambda$, namely for $\lambda=0.0,0.25,0.5,0.75,1.0$. Convergence
was checked by comparison with values obtained using the trapezoidal
rule. Since the dependence of the estimator on the parameter $\lambda$
is almost linear, the difference between the two results remained
well under the statistical error. Using the SCC-DFTB potential, the
dependence on $\lambda$ was less smooth. As a result, nine equidistant
values of $\lambda$ were needed to achieve similar convergence. In
one case, as many as 17 values of $\lambda$ had to be used. 

\begin{table*}
\begin{tabular}{lcccc}
\hline 
&
1$^{\text{st}}$ step $\Delta F^{\mathrm{red}}$ &
$\Delta\Delta F^{\mathrm{anharm}}$ &
 2$^{\text{nd}}$ step $\Delta F^{\mathrm{red}}$ &
$\Delta\Delta F^{\mathrm{anharm}}$ \tabularnewline
\hline 
\multicolumn{5}{l}{(3Z)-(5,5,5-$^{\text{2}}$H$_{\text{3}}$)penta-1,3-diene (\textbf{1-}\textbf{\footnotesize 5,5,5}\textbf{-}\textbf{\textit{d}}\textbf{$_{\text{3}}$})}\tabularnewline
\hline 
AM1 (PIMD)&
0.0395&
-0.0041$\pm$0.0009&
-0.0154&
0.0022$\pm$0.0007\tabularnewline
SCC-DFTB (PIMD)&
0.1245&
-0.0039$\pm$0.0007&
-0.0616&
0.0026$\pm$0.0005\tabularnewline
B98 (HA) + $\Delta\Delta F_{\mathrm{AM1}}^{\mathrm{anharm}}$&
0.0587&
&
-0.0248&
\tabularnewline
MP2 (HA) + $\Delta\Delta F_{\mathrm{AM1}}^{\mathrm{anharm}}$&
0.0770&
&
-0.0338&
\tabularnewline
\hline 
\multicolumn{5}{l}{(3Z)-(1,1-$^{\text{2}}$H$_{2}$)penta-1,3-diene (\textbf{1-}\textbf{\footnotesize 1,1}\textbf{-}\textbf{\textit{d}}\textbf{$_{\text{2}}$})}\tabularnewline
\hline
AM1 (PIMD)&
-0.0283&
0.0063$\pm$0.0009&
0.0191&
-0.0023$\pm$0.0007\tabularnewline
SCC-DFTB (PIMD)&
-0.1142&
0.0049$\pm$0.0006&
0.0610&
-0.0017$\pm$0.0005\tabularnewline
B98 (HA) + $\Delta\Delta F_{\mathrm{AM1}}^{\mathrm{anharm}}$&
-0.0466&
&
0.0282&
\tabularnewline
MP2 (HA) + $\Delta\Delta F_{\mathrm{AM1}}^{\mathrm{anharm}}$&
-0.0645&
&
0.0372&
\tabularnewline
\hline
\end{tabular}
\caption{Reduced  free energies \textsf{$\Delta F^{\mathrm{red}}$} $\left[\mathrm{kcal\cdot mol^{-1}}\right]$
and anharmonicity corrections $\Delta\Delta F^{\mathrm{anharm}}$
$\left[\mathrm{kcal\cdot mol^{-1}}\right]$ of {[}1,5] hydrogen shift
reactions in \textbf{1-}\textbf{\footnotesize 5,5,5}\textbf{-}\textbf{\textit{d}}\textbf{$_{\text{3}}$}
and in \textbf{1-}\textbf{\footnotesize 1,1}\textbf{-}\textbf{\textit{d}}\textbf{$_{\text{2}}$}
at 478.45 $\mathrm{K}$. For AM1 and SCC-DFTB methods, values calculated
by PIMD are listed followed by the anharmonicity correction obtained
as the difference of the PIMD and HA value. For B98 and MP2 methods,
HA values corrected by the AM1 anharmonicity correction are listed.
The first reaction step in the case of tri-deuterated compound leads
from \textbf{1-}\textbf{\footnotesize 5,5,5}\textbf{-}\textbf{\textit{d}}\textbf{$_{\text{3}}$}
to \textbf{1-}\textbf{\footnotesize 1,1,5}\textbf{-}\textbf{\textit{d}}\textbf{$_{\text{3}}$},
which is also the reactant of the second reaction step leading to
\textbf{1-}\textbf{\footnotesize 1,5,5}\textbf{-}\textbf{\textit{d}}\textbf{$_{\text{3}}$}.
In case of di-deuterated compound the sequence is: \textbf{1-}\textbf{\footnotesize 1,1}\textbf{-}\textbf{\textit{d}}\textbf{$_{\text{2}}$},
\textbf{1-}\textbf{\footnotesize 5,5}\textbf{-}\textbf{\textit{d}}\textbf{$_{\text{2}}$}
and \textbf{1-}\textbf{\footnotesize 1,5}\textbf{-}\textbf{\textit{d}}\textbf{$_{\text{2}}$}.
\label{tab:F_pentadiene}}
\end{table*}

\subsection*{Amber 10 implementation}

The PIMD calculations were performed using Amber 10.\cite{amber10}
The part of Amber 10 code, which computes the derivative $dF\left(\lambda\right)/d\lambda$
with respect to the mass was implemented by one of us and can be invoked
by setting ITIMASS variable in the input file. Several posible ways
to compute the derivative are obtained by combining one of two implementations
of PIMD in \texttt{sander (}either the multisander implementation
or the LES implementation) with either the Nos\'{e}-Hoover chains
of thermostats or the Langevin thermostat, with the normal mode or
{}``primitive'' PIMD, and with the TE or GVE. Calculating the value
of $dF\left(\lambda\right)/d\lambda$ for compound \textbf{1-}\textbf{\footnotesize 5,5,5}\textbf{-}\textbf{\textit{d}}\textbf{$_{\text{3}}$}
using GAFF force field, at $\lambda=0$, $T=478.45$ $\mathrm{K}$,
and for several values of $P$, we confirmed that all twelve possible
combinations give the same result. For example, Fig. \ref{fig:GVE_TE_values}
shows the agreement of the GVE and TE. Nevertheless, the twelve methods
differ by RMSEs of $dF\left(\lambda\right)/d\lambda$ (due to different
statistical errors of estimators and correlation lengths) and by computational
costs. As expected, the most important at higher values of $P$ is
the difference of statistical errors of GVE and TE. Figure \ref{fig:GVE_TE_errors}
compares the dependence of the RMSE of the GVE and TE on the number
$P$ of imaginary time slices. For $P=64$, the converged simulation
with GVE is approximately 100 times faster than with TE. Less significant
differences of RMSEs are due to differences in correlation lengths.
As expected, for primitive PIMD with Langevin thermostat the correlation
length depends strongly on collision frequency $\gamma$ of the thermostat.
The correlation length is approximately 450-500 $\mathrm{fs}$ for
$\gamma=0.3$ $\mathrm{ps^{-1}}$, falling down quickly to 120-150
$\mathrm{fs}$ for $\gamma=3$ $\mathrm{ps^{-1}}$ and to 5 $\mathrm{fs}$
for $\gamma=300$ $\mathrm{ps^{-1}}$, then rising slowly again. This
can be compared to correlation length of 10-20 $\mathrm{fs}$ of the
primitive PIMD thermostated by Nos\'{e}-Hoover chains of 4 thermostats
per degree of freedom. A smaller difference can be found between correlation
lengths of the normal mode (3-8 $\mathrm{fs}$) and primitive PIMD
(10-20 $\mathrm{fs}$).

\subsection*{Used software}

All PIMD calculations were performed in Amber 10.\cite{amber10} All
MP2 and B98 calculations as well as AM1 and PM3 semiempirical calculations
in the HA were done in Gaussian 03 revision E01.\cite{g03} DFTB and
SCC-DFTB calculations in the HA used the DFTB+ code, version 1.0.1.\cite{AradiB._jp070186p}
SCC-DFTB harmonic frequencies were computed numerically using analytical
gradients provided by the DFTB+ code. The step size for numerical
differentiation was set equal to 0.01 $\textrm{\AA}$. This value
was also used by Kr\"{u}ger \textit{et al.}\cite{kruger:114110}
in their study validating the SCC-DFTB method and their frequencies
differed from purely analytical frequencies of Witek \textit{et al.}\cite{Witek2004}
by at most 10 $\mathrm{cm^{-1}}$. To diagonalize the resulting numerical
Hessian, we used the \texttt{formchk} utility included in the Gaussian
program package.

\section{Results\label{sec:Results}}

\subsection{(3\textit{Z})-penta-1,3-diene}

(3\textit{Z})-penta-1,3-diene (compound \textbf{1}) is the simplest
of examined molecules. Its non-deuterated isotopolog has three distinguishable
minima: the s-\textit{trans} conformer, which is the global minimum
and two s-\textit{cis} conformers related by  mirror symmetry.  Strictly
speaking, in pentadiene the s-\textit{cis} species have actually \textit{gauche}
conformations due to sterical constraints. In their original experiments,
Roth and K\"onig studied two isotopologs, \textbf{1-}\textbf{\footnotesize 5,5,5}\textbf{-}\textbf{\textit{d}}\textbf{$_{\text{3}}$}
and \textbf{1-}\textbf{\footnotesize 1,1}\textbf{-}\textbf{\textit{d}}\textbf{$_{\text{2}}$}.
The EIEs of both isotopologs were computed using the PIMD methodology
of Sec. \ref{sec:Method} at 478.45 $\mathrm{K}$. Resulting reduced
 free energies are listed in Table \ref{tab:F_pentadiene}. %
 Note that the anharmonicity correction is very similar for AM1 and
SCC-DFTB methods, even though the main part of \textsf{$\Delta F^{\mathrm{red}}$}
obtained in the HA (\textsf{$\Delta F_{\mathrm{HA}}^{\mathrm{red}}$)}
substantially differs for the two methods. This indicates that the
corrections are fairly reliable in this case and can be used to correct
 results of higher level methods obtained in the HA. The anharmonicity
correction is as large as 20~\% of the final value of the reduced
 free energy. Unfortunately, the difference between MP2 and B98 in
the HA is still approximately four times larger than the anharmonicity
correction. 

\begin{table}[H]
\begin{tabular}{lccc}
\hline 
&
\textbf{1-}\textbf{\footnotesize 5,5,5}\textbf{-}\textbf{\textit{d}}\textbf{$_{\text{3}}$}&
\textbf{1-}\textbf{\footnotesize 1,1,5}\textbf{-}\textbf{\textit{d}}\textbf{$_{\text{3}}$}&
\textbf{1-}\textbf{\footnotesize 1,5,5}\textbf{-}\textbf{\textit{d}}\textbf{$_{\text{3}}$}\tabularnewline
\hline
AM1 (PIMD)&
0.103&
0.296&
0.601\tabularnewline
SCC-DFTB (PIMD)&
0.108&
0.285&
0606\tabularnewline
B98 (HA) + $\Delta\Delta F_{\mathrm{AM1}}^{\mathrm{anharm}}$&
0.104&
0.293&
0.602\tabularnewline
MP2 (HA) + $\Delta\Delta F_{\mathrm{AM1}}^{\mathrm{anharm}}$&
0.105&
0.291&
0.604\tabularnewline
\hline 
&
\textbf{1-}\textbf{\footnotesize 1,1}\textbf{-}\textbf{\textit{d}}\textbf{$_{\text{2}}$}&
\textbf{1-}\textbf{\footnotesize 5,5}\textbf{-}\textbf{\textit{d}}\textbf{$_{\text{2}}$}&
\textbf{1-}\textbf{\footnotesize 1,5}\textbf{-}\textbf{\textit{d}}\textbf{$_{\text{2}}$}\tabularnewline
\hline
AM1 (PIMD)&
0.099&
0.305&
0.597\tabularnewline
SCC-DFTB (PIMD)&
0.093&
0.315&
0.591\tabularnewline
B98 (HA) + $\Delta\Delta F_{\mathrm{AM1}}^{\mathrm{anharm}}$&
0.097&
0.307&
0.596\tabularnewline
MP2 (HA) + $\Delta\Delta F_{\mathrm{AM1}}^{\mathrm{anharm}}$&
0.096&
0.309&
0.595\tabularnewline
\hline
\end{tabular}

\caption{Equilibrium ratios of {[}1,5] hydrogen shift reactions of \textbf{1-}\textbf{\footnotesize 5,5,5}\textbf{-}\textbf{\textit{d}}\textbf{$_{\text{3}}$}
and \textbf{1-}\textbf{\footnotesize 1,1}\textbf{-}\textbf{\textit{d}}\textbf{$_{\text{2}}$}
at 478.45 $\mathrm{K}$. \label{tab:eq_ratios_pentadiene}}
\end{table}

\begin{table*}
\begin{tabular}{lcccc}
\hline 
&
1$^{\text{st}}$ step $\Delta F^{\mathrm{red}}$&
$\Delta\Delta F^{\mathrm{anharm}}$ &
2$^{\text{nd}}$ step $\Delta F^{\mathrm{red}}$ &
$\Delta\Delta F^{\mathrm{anharm}}$ \tabularnewline
\hline
AM1 (PIMD)&
-0.0337&
0.0102$\pm$0.0013&
0.0230&
-0.0036$\pm$0.0010\tabularnewline
B98 (HA) + $\Delta\Delta F_{\mathrm{AM1}}^{\mathrm{anharm}}$&
-0.0496&
&
0.0302&
\tabularnewline
MP2 (HA) + $\Delta\Delta F_{\mathrm{AM1}}^{\mathrm{anharm}}$&
-0.0674&
&
0.0392&
\tabularnewline
\hline
\end{tabular}

\caption{Reduced  free energies \textsf{$\Delta F^{\mathrm{red}}$} $\left[\mathrm{kcal\cdot mol^{-1}}\right]$
and anharmonicity corrections $\Delta\Delta F^{\mathrm{anharm}}$
$\left[\mathrm{kcal\cdot mol^{-1}}\right]$ of the {[}1,5] hydrogen
shift reaction in 2-methyl-10-(10,10-$^{\text{2}}$H$_{2}$)methylenebicyclo{[}4.4.0]dec-1-ene
(compound \textbf{2}) at 441.05 $\mathrm{K}$. For AM1 method, values
calculated by PIMD are listed followed by the anharmonicity correction
obtained as the difference of PIMD and HA values. For B98 and MP2
methods, only HA values corrected by the AM1 anharmonicity correction
are listed. The first reaction step leads from \textbf{2-}\textbf{\footnotesize 1,1}\textbf{-}\textbf{\textit{d}}\textbf{$_{\text{2}}$}
to \textbf{2-}\textbf{\footnotesize 5,5}\textbf{-}\textbf{\textit{d}}\textbf{$_{\text{2}}$},
which is also the reactant of the second reaction step leading to
\textbf{2-}\textbf{\footnotesize 1,5}\textbf{-}\textbf{\textit{d}}\textbf{$_{\text{2}}$}.
\label{tab:F_bicyclo}}
\end{table*}
Reduced free energies suggest a general preference valid for both
tri- and di-deuterio species: Namely, deuterium, compared to hydrogen,
prefers sp$^{\text{3}}$ carbon of the methyl group to sp$^{\text{2}}$
carbon of the vinyl group. This was also observed experimentally.\cite{SunkoDionisE._ja00725a026,Barborak1971,Gajewski1979}
As can be seen in the table, the preference is present already in
the HA. At first sight this preference can be counter-intuitive: If
we confine ourselves only to the most energetical C-H (or C-D) bond
stretching modes and suppose that force constants do not  change significantly
upon substitution with deuterium, the deuterium should prefer the
stiffest bonds. This is because more energy can be gained by substituting
the stiffer sp$^{2}$ C-H bonds, assuming approximately the same change
in reduced mass after substitution. (Recall that the energy of a vibrational
mode is proportional to $\sqrt{k/\mu}$ where $k$ is the force constant
and $\mu$ the reduced mass.) Considering the stretching modes only,
this would be the case for isotopologs of compound \textbf{1}. Nevertheless,
taking into account also the bending and torsional vibrational modes,
gains on the sp$^{\text{3}}$ C-H bond side will dominate. (This {}``counter-intuitive''
preference of heavier isotope in {}``softer'' bonds is quite common.
For examples see the inverse H/D equilibrium isotope effect in oxidative
addition reactions of H$_{\text{2}}$ to transition metal complexes,\cite{Hascall1999,Bender1995,Abu-Hasanayn1993,Rabinovich1993}
or the inverse $^{\text{16}}$O/$^{\text{18}}$O isotope effect in
metal mediated oxygen activation reaction.\cite{Smirnov2009}) As
already stressed, final equilibrium concentrations are determined
mainly by the symmetry factors and the aforementioned deuterium sp$^{\text{3}}$
to sp$^{\text{2}}$ preference manifests itself only in a small modification
of the symmetry determined rational ratios, as seen in Table \ref{tab:eq_ratios_pentadiene}.

\subsubsection*{Temperature dependence of the reduced free energy}

\begin{figure}
\includegraphics{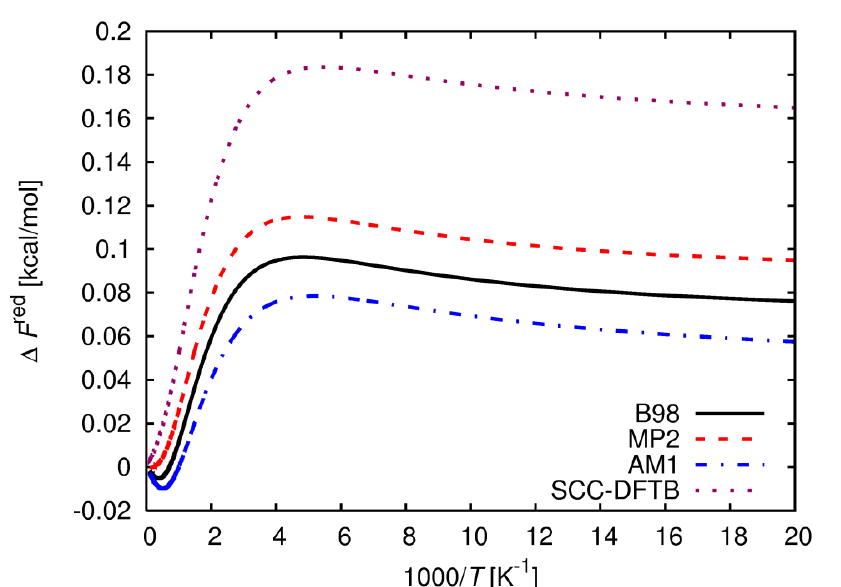}

\caption{Reduced reaction free energies of the first step of {[}1,5] hydrogen
shift reaction in (3\textit{Z})-(5,5,5-$^{\text{2}}$H$_{\text{3}}$)penta-1,3-diene
(\textbf{1-}\textbf{\footnotesize 5,5,5}\textbf{-}\textbf{\textit{d}}\textbf{$_{\text{3}}$})
calculated in the HA as the Boltzmann average of all s-\textit{trans}
and s-\textit{cis} isomers.\label{fig:harmonic_freq_semiempirical_test}}
\end{figure}

Temperature dependence of the reduced  free energy for the first reaction
step of hydrogen shift in \textbf{1-}\textbf{\footnotesize 5,5,5}\textbf{-}\textbf{\textit{d}}\textbf{$_{\text{3}}$}
 is depicted in Fig. \ref{fig:harmonic_freq_semiempirical_test}.
Analogous temperature dependence for all other studied reactions is
very similar to the dependence in this figure. At very low temperature,
the reduced reaction free energy approaches the difference of ZPEs
in accordance with Eq. \eqref{eq:EIE_low_temp}. Absolute value of
the reduced  free energy is maximal at temperatures around 200 $\mathrm{K}$.
At temperatures around 400-500 $\mathrm{K}$, where most measurements
took place, the value of \textsf{$\Delta F^{\mathrm{red}}$} is (by
chance) close to the difference of ZPEs. At high temperatures, \textsf{$\Delta F^{\mathrm{red}}$}
goes to zero in accordance with the high temperature limit \eqref{eq:EIE_class}
discussed above, which is valid also in the HA as can be shown using
the Teller-Redlich theorem.\cite{Angus1935,Redlich1935} The temperature
dependence of the anharmonicity correction was examined using the
AM1 semiempirical method at temperatures 200 $\mathrm{K}$, 478.45
$\mathrm{K}$ and 1000 $\mathrm{K}$. The correction  for the first
reaction step of \textbf{1-}\textbf{\footnotesize 5,5,5}\textbf{-}\textbf{\textit{d}}\textbf{$_{\text{3}}$}
is equal to $-0.0042$ $\pm$ $0.0009$, $-0.0041$ $\pm$ $0.0009$
and $-0.0035$ $\pm$ $0.0008$ $\mathrm{kcal\cdot mol^{-1}}$, respectively.
Therefore, taking into account statistical errors, the correction
stays approximately constant over the wide temperature range. Since
the value of \textsf{$\Delta F^{\mathrm{red}}$} is decreasing in
the region from 200 to 1000 $\mathrm{K}$, the relative importance
of the correction is increasing. %

\subsection{2-methyl-10-methylenebicyclo{[}4.4.0]dec-1-ene }

\begin{table}[H]
\begin{tabular}{lccc}
\hline 
&
\textbf{2-}\textbf{\footnotesize 1,1}\textbf{-}\textbf{\textit{d}}\textbf{$_{\text{2}}$}&
\textbf{2-}\textbf{\footnotesize 5,5}\textbf{-}\textbf{\textit{d}}\textbf{$_{\text{2}}$}&
\textbf{2-}\textbf{\footnotesize 1,5}\textbf{-}\textbf{\textit{d}}\textbf{$_{\text{2}}$}\tabularnewline
\hline
AM1 (PIMD)&
0.098&
0.306&
0.595\tabularnewline
B98 (HA) + $\Delta\Delta F_{\mathrm{AM1}}^{\mathrm{anharm}}$&
0.097&
0.308&
0.595\tabularnewline
MP2 (HA) + $\Delta\Delta F_{\mathrm{AM1}}^{\mathrm{anharm}}$&
0.096&
0.310&
0.594\tabularnewline
experimental (series 1)\footnotemark[1]&
0.108&
0.328&
0.564\tabularnewline
experimental (series 2)\footnotemark[1]&
0.114&
0.314&
0.572\tabularnewline
\hline
\end{tabular}

\footnotetext[1]{Ref. \onlinecite{Doering2006}}

\caption{Equilibrium ratio of the {[}1,5] hydrogen shift reaction in dideuterated
compound \textbf{2} at 441.05 $\mathrm{K}$. Experimental series 1
and 2 were obtained by two different methods of analysis of the NMR
spectrum.\cite{Doering2006}}

\label{tab:eq_ratios_bicyclo}
\end{table}

\begin{table*}
\begin{tabular}{lcccc}
\hline 
&
1$^{\text{st}}$ step $\Delta F^{\mathrm{red}}$&
$\Delta\Delta F^{\mathrm{anharm}}$ &
2$^{\text{nd}}$ step $\Delta F^{\mathrm{red}}$ &
$\Delta\Delta F^{\mathrm{anharm}}$ \tabularnewline
\hline
AM1 (PIMD)&
-0.0439&
0.0160$\pm$0.0015&
0.0265&
-0.0080$\pm$0.0011\tabularnewline
\multicolumn{1}{l}{B98 (HA) + $\Delta\Delta F_{\mathrm{AM1}}^{\mathrm{anharm}}$}&
-0.0673&
&
0.0376&
\tabularnewline
\hline
\end{tabular}

\caption{Reduced free energies \textsf{$\Delta F^{\mathrm{red}}$} $\left[\mathrm{kcal\cdot mol^{-1}}\right]$
and anharmonicity corrections $\Delta\Delta F^{\mathrm{anharm}}$
$\left[\mathrm{kcal\cdot mol^{-1}}\right]$ of the {[}1,5] hydrogen
shift reaction in 2,4,6,7,9-pentamethyl-5-(5,5-$^{\text{2}}$H$_{2}$)methylene-11,11a-dihydro-12\textit{H}-naphthacene
(compound \textbf{3}) at 441.05 $\mathrm{K}$. For AM1 method, values
calculated by PIMD are listed followed by the anharmonicity correction
obtained as the difference of PIMD and HA values. For B98, HA values
corrected by the AM1 anharmonicity corrections are listed. The first
reaction step leads from \textbf{2-}\textbf{\footnotesize 1,1}\textbf{-}\textbf{\textit{d}}\textbf{$_{\text{2}}$}
to \textbf{2-}\textbf{\footnotesize 5,5}\textbf{-}\textbf{\textit{d}}\textbf{$_{\text{2}}$},
which is also the reactant of the second reaction step leading to
\textbf{2-}\textbf{\footnotesize 1,5}\textbf{-}\textbf{\textit{d}}\textbf{$_{\text{2}}$}.
 \label{tab:F_naphthacene}}
\end{table*}

Compound \textbf{2-}\textbf{\footnotesize 1,1}\textbf{-}\textbf{\textit{d}}\textbf{$_{\text{2}}$}
was used relatively recently by Doering and Zhao\cite{Doering2006}
to refine the original results of Roth and K\"onig. Doering and Zhao
reported the equilibrium concentrations at three temperatures, from
which we have chosen the lowest, $T=441.05$ $\mathrm{K}$. Since
AM1 and SCC-DFTB methods gave similar anharmonicity corrections for
compound \textbf{1}, we used only the AM1 method in this case. Four
minima were found.  Energy differences between the global minimum
and local minima at the B98/6-311+(2df,p) level of theory are 4.8,
7.5 and 9.5 $\mathrm{kcal\cdot mol^{-1}}$. Resulting \textsf{\footnotesize }\textsf{$\Delta F^{\mathrm{red}}$}
calculated according to Eq. \eqref{eq:F_red_HA} are listed in Table
\ref{tab:F_bicyclo}. As expected, they are similar to \textsf{$\Delta F^{\mathrm{red}}$}
of \textbf{1-}\textbf{\footnotesize 1,1}\textbf{-}\textbf{\textit{d}}\textbf{$_{\text{2}}$}.
Anharmonicity corrections changed more and they are about 60~\% higher
in the absolute value. %

Calculated equilibrium ratios are listed in Table \ref{tab:eq_ratios_bicyclo}
together with experimental ratios reported by Doering and Zhao. Theoretical
and experimental ratios differ substantially, which suggests that
side reactions suspected by Doering and Zhao had indeed occurred and
influenced the accuracy of results of their study.

\subsection{2,4,6,7,9-pentamethyl-5-methylene-11,11a-dihydro-12\textit{H}-naphthacene}

In order to suppress unwanted side reactions suspected for compound
\textbf{2},\cite{Doering2006} Doering and Keliher further developed
the model compound into \textbf{3-}\textbf{\footnotesize 1,1}\textbf{-}\textbf{\textit{d}}\textbf{$_{\text{2}}$}
by adding methyl substituted aromatic rings on both sides of the cyclic
part.\cite{Doering2007} With this compound they obtained the same
equilibrium ratios for all temperatures they had examined.\cite{Doering2007}
Because of this and because the temperature 441.05 $\mathrm{K}$ we
have chosen for our analysis of compound \textbf{2} differs from one
of temperatures used in Ref. \onlinecite{Doering2007} by less than
2 $\mathrm{K}$, we have decided to use this temperature also for
compound \textbf{3}. In the HA, only the B98 density functional method
was used, due to a considerable size of the molecule. Because of the
increased rigidity imposed by aromatic rings on the sides of the original
bi-cyclic compound, only three distinct minima were found. (We neglected
several possible orientations of two methyl groups distant from the
reaction site, which hardly affect the final result.) At the B98/6-311+G(2df,p)
level, the local minima have energies 1.3 and 5.6 $\mathrm{kcal\cdot mol^{-1}}$
above the global minimum. The AM1 method gives the opposite order
of the first and second lowest minima. Reduced  free energies and
anharmonicity corrections obtained using the AM1 method are listed
in Table \ref{tab:F_naphthacene}.

Values of both \textsf{$\Delta F^{\mathrm{red}}$} and the anharmonicity
corrections are again qualitatively similar (but higher in absolute
values) to those of the smaller and less strained compound \textbf{2}.
Here, values of anharmonicity corrections reached approximately 30~\%
of the values of reduced  free energies of both reaction steps. Resulting
equilibrium ratios together with their experimental values can be
seen in Table \ref{tab:eq_ratios_naphthacene}. Agreement of the theoretical
prediction with the experimental result is very good. An uncorrected
B98 HA value is also included in Table \ref{tab:eq_ratios_naphthacene},
to demonstrate how the anharmonicity correction modifies the HA equilibrium
ratio. Surprisingly, the direct AM1 PIMD calculation is closest to
the experimental value, but this ought to be ascribed to a fortunate
coincidence, considering the aforementioned accuracy of electronic
structure methods used in our study. 

\begin{table}
\begin{tabular}{lccc}
\hline 
&
\textbf{3-}\textbf{\footnotesize 1,1}\textbf{-}\textbf{\textit{d}}\textbf{$_{\text{2}}$}&
\textbf{3-}\textbf{\footnotesize 5,5}\textbf{-}\textbf{\textit{d}}\textbf{$_{\text{2}}$}&
\textbf{3-}\textbf{\footnotesize 1,5}\textbf{-}\textbf{\textit{d}}\textbf{$_{\text{2}}$}\tabularnewline
\hline
AM1 (PIMD)&
0.098&
0.308&
0.595\tabularnewline
B98 (HA)&
0.095&
0.312&
0.593\tabularnewline
B98 (HA) + $\Delta\Delta F_{\mathrm{AM1}}^{\mathrm{anharm}}$&
0.096&
0.310&
0.594\tabularnewline
experimental\footnotemark[1]&
0.098&
0.308&
0.594\tabularnewline
\hline
\end{tabular}

\footnotetext[1]{Ref. \onlinecite{Doering2007}}

\caption{Equilibrium ratios of the {[}1,5] hydrogen shift reaction in compound
\textbf{3} at 441.05 $\mathrm{K}$. }

\label{tab:eq_ratios_naphthacene}
\end{table}

\section{Discussion and conclusions\label{sec:Discussion-and-conclusions}}

To conclude, the combination of higher level methods in the HA with
PIMD using semiempirical methods for the rigorous treatment of effects
beyond the HA proved to be a viable method for accurate calculations
of EIEs. Using the generalized virial estimator for the derivative
of the free energy with respect to the mass we were able to obtain
accurate results at lower temperatures in reasonable time ($\sim60$
times faster than with thermodynamic estimator), since the statistical
error is independent on the number of imaginary time slices. Two semiempirical
methods, AM1 and SCC-DFTB, were used for calculation of the anharmonicity
correction, both giving very similar results. Calculations showed,
that the anharmonicity effects account up to 30 \% of the final value
of the reduced free energy of considered reactions. The anharmonicity
correction always decreases the absolute value of the reduced reaction
free energy. This is consistent with the qualitative picture in which
higher vibrational frequencies of hydrogens are lowered by the anharmonicity
of a potential more than lower frequencies of deuteriums. This in
turn is due to a higher amplitude of vibrations of lighter hydrogen
atoms. The lower difference between frequencies of unsubstituted and
deuterated species results in the lower absolute value of the reduced
reaction free energy. Unfortunately, the inaccuracy of the \textit{ab
initio} electronic structure methods used in our study is still of
at least the same order as the anharmonicity corrections.

For isotopologs of compound \textbf{1}, we predicted equilibrium ratios
and free energies of the {[}1,5] sigmatropic hydrogen shift reaction.
A comparison with experimental results was not possible due to a low
precision of the original measurement. For compound \textbf{2}, the
disagreement between theoretical and experimental data supports the
suspicion by authors of the measurement that the accuracy of their
results was compromised by dimerization side reactions. On the other
hand, the agreement of theoretically calculated ratios with experimental
observations in the case of compound \textbf{3} suggests that the
isolation of the {[}1,5] hydrogen shift reaction from disturbing influences
was successfully achieved and the observed EIE and KIE can be considered
reliable.

\begin{acknowledgments}
We acknowledge the start-up funding provided by EPFL. We express our
gratitude to Daniel Jana for his assistance with SCC-DFTB calculations,
to H. Witek for providing us the set of frequency optimized SCC-DFTB
parameters, and to J. J. P. Stewart for the MOPAC2007 code used for
testing the PM6 method.
\end{acknowledgments}

\appendix

\section{Examination of semiempirical methods used in PIMD simulations}

To determine the suitability of semiempirical methods used in our
PIMD calculations, we first compared values of the EIE in the HA.
The B98 and MP2 methods served as a reference. As can be seen from
Fig. 6 in the paper, which shows the temperature dependence of $\Delta F^{\mathrm{red}}$
of the first step of deuterium transfer reaction in \textbf{1-}\textbf{\footnotesize 5,5,5}\textbf{-}\textbf{\textit{d}}\textbf{$_{\text{3}}$},
the difference between B98 and MP2 at  temperatures below 500 $\mathrm{K}$
is close to 0.02 kcal/mol. So is the difference between AM1 and B98
methods. On the other hand, the SCC-DFTB method clearly overestimates
the extent of EIE compared to both higher level methods. A very similar
trend was observed in all examined reactions. Other semiempirical
methods tested were PM3\cite{Stewart1989} and SCC-DFTB-MWM, which
are not included in Fig. 6 in the paper for clarity. The PM3 method
overestimates the EIE similarly to SCC-DFTB, whereas the SCC-DFTB-MWM
curve is somewhat closer to \textit{ab initio} curves than the SCC-DFTB
one. To conclude, from this point of view AM1 is the preferred semiempirical
method.

\begin{figure}
\includegraphics{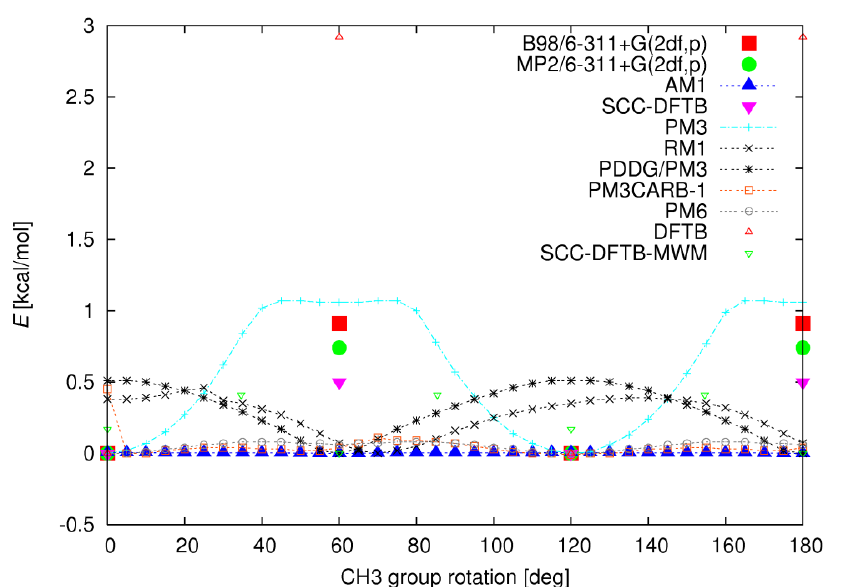}

\caption{Relaxed potential energy scan of the methyl rotation in s-\textit{trans}
(3\textit{Z})-penta-1,3-diene (compound \textbf{1}). For MP2, B98,
DFTB, SCC-DFTB and SCC-DFTB-MWM only positions and potential energies
of minima and maxima are indicated. \label{fig:methyl_scan}}
\end{figure}

\begin{figure}
\includegraphics{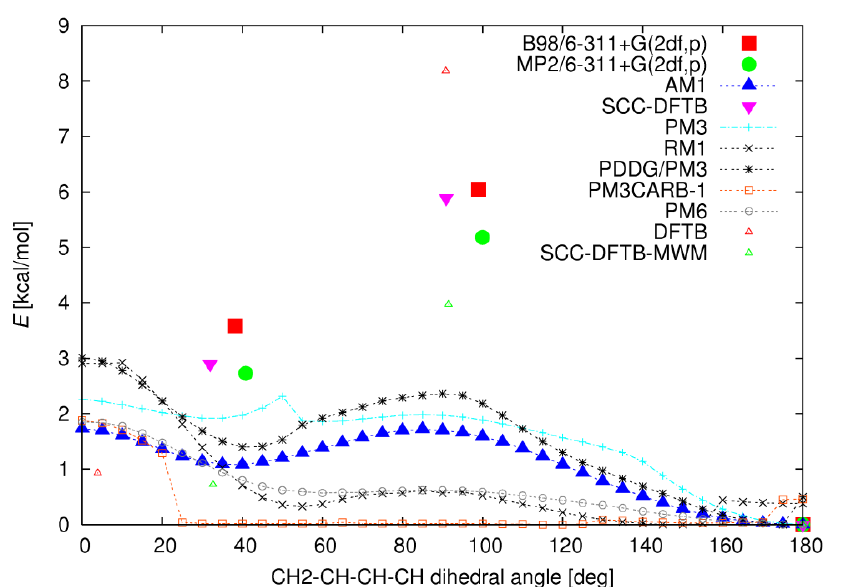}

\caption{Relaxed potential energy scan of the vinyl group rotation in s-\textit{trans}
(3\textit{Z})-penta-1,3-diene (compound \textbf{1}). For MP2, B98,
DFTB, SCC-DFTB and SCC-DFTB-MWM only positions and potential energies
of minima and transition states are indicated.\label{fig:vinyl_scan}}
\end{figure}

During simulations, the pentadiene molecule often passes two potential
energy barriers. These are the barrier for the hindered rotation of
the methyl group and the barrier for the rotation of the vinyl group,
which connects s-\textit{trans} and s-\textit{cis} conformations.
Relaxed potential energy scans of these two motions were employed
as the second criterion to assess the relevancy of semiempirical methods.
Methods tested were MP2, B98, AM1, SCC-DFTB, SCC-DFTB-MWM, PM3, RM1,\cite{Rocha2006}
PM3CARB-1,\cite{McNamara2004} PDDG/PM3,\cite{Repasky2002} and PM6.\cite{Stewart2007}
Potential surface scans with PM3CARB1, RM1, PM3/PDDG methods were
calculated using the public domain code MOPAC 6. PM6 potential surface
scans were performed in MOPAC 2007.\cite{mopac2007} Results for the
methyl group rotation are shown in Fig. \ref{fig:methyl_scan}. The
height of the AM1 barrier  is only 0.005 $\mathrm{kcal\cdot mol^{-1}}$.
Moreover, positions of minima do not agree with B98 and MP2. On the
other hand, the SCC-DFTB method matches higher level methods closely.
From other semiempirical methods PM3 performs best in this aspect.
The height of the barrier is relatively well reproduced also by PDDG/PM3,
RM1 and SCC-DFTB-MWM methods, but positions of extrema of the potential
energy surface are incorrect.%
Figure \ref{fig:vinyl_scan} shows potential energy scans of the s-\textit{trans}/s-\textit{cis}
rotation of the vinyl group. Again, the SCC-DFTB method matches higher
level methods closely. All other methods (with the exception of DFTB)
give too low barrier heights as well as too low energy differences
between s-\textit{trans} and s-\textit{cis} conformations. Also note
that the potential energy surfaces of PM3 and related methods (PDDG/PM3
and PM3CARB-1) are not smooth in the gauche region. This peculiarity
of the PM3 potential surface can be seen also in the potential surface
scan performed by Liu \textit{et al.}\cite{Liu1993} %
Based on these results we concluded that none of the semiempirical
methods except for SCC-DFTB is able to sufficiently improve the AM1
potential energy surface. Whereas the frequency optimized variant
of the SCC-DFTB method (SCC-DFTB-MWM) improves the EIE in the HA,
it does not retain the SCC-DFTB accuracy in the potential surface
scans. Hence we decided to use the AM1 and SCC-DFTB potentials for
PIMD calculations. To conclude, AM1 performs very well in HA, but
it cannot properly describe potential surfaces of the two rotational
motions realized during simulations. On the other hand,  SCC-DFTB
gives worse results in HA,  but it reproduces both barriers very well.

\bibliographystyle{apsrev}

\begin{thebibliography}{68}
\expandafter\ifx\csname natexlab\endcsname\relax\def\natexlab#1{#1}\fi
\expandafter\ifx\csname bibnamefont\endcsname\relax
  \def\bibnamefont#1{#1}\fi
\expandafter\ifx\csname bibfnamefont\endcsname\relax
  \def\bibfnamefont#1{#1}\fi
\expandafter\ifx\csname citenamefont\endcsname\relax
  \def\citenamefont#1{#1}\fi
\expandafter\ifx\csname url\endcsname\relax
  \def\url#1{\texttt{#1}}\fi
\expandafter\ifx\csname urlprefix\endcsname\relax\def\urlprefix{URL }\fi
\providecommand{\bibinfo}[2]{#2}
\providecommand{\eprint}[2][]{\url{#2}}

\bibitem[{\citenamefont{Alston et~al.}(1996)\citenamefont{Alston, Haley,
  Kanski, Murray, and Pranata}}]{Alston1996}
\bibinfo{author}{\bibfnamefont{W.~C.} \bibnamefont{Alston}},
  \bibinfo{author}{\bibfnamefont{K.}~\bibnamefont{Haley}},
  \bibinfo{author}{\bibfnamefont{R.}~\bibnamefont{Kanski}},
  \bibinfo{author}{\bibfnamefont{C.~J.} \bibnamefont{Murray}},
  \bibnamefont{and} \bibinfo{author}{\bibfnamefont{J.}~\bibnamefont{Pranata}},
  \bibinfo{journal}{J. Am. Chem. Soc.} \textbf{\bibinfo{volume}{118}},
  \bibinfo{pages}{6562} (\bibinfo{year}{1996}).

\bibitem[{\citenamefont{Anbar et~al.}(2005)\citenamefont{Anbar, Jarzecki, and
  Spiro}}]{Anbar2005825}
\bibinfo{author}{\bibfnamefont{A.}~\bibnamefont{Anbar}},
  \bibinfo{author}{\bibfnamefont{A.}~\bibnamefont{Jarzecki}}, \bibnamefont{and}
  \bibinfo{author}{\bibfnamefont{T.}~\bibnamefont{Spiro}},
  \bibinfo{journal}{Geochim. Cosmochim. Acta} \textbf{\bibinfo{volume}{69}},
  \bibinfo{pages}{825 } (\bibinfo{year}{2005}).

\bibitem[{\citenamefont{Gawlita et~al.}(1994)\citenamefont{Gawlita, Anderson,
  and Paneth}}]{Gawlita1994}
\bibinfo{author}{\bibfnamefont{E.}~\bibnamefont{Gawlita}},
  \bibinfo{author}{\bibfnamefont{V.~E.} \bibnamefont{Anderson}},
  \bibnamefont{and} \bibinfo{author}{\bibfnamefont{P.}~\bibnamefont{Paneth}},
  \bibinfo{journal}{Eur. Biophys. J.} \textbf{\bibinfo{volume}{23}},
  \bibinfo{pages}{353} (\bibinfo{year}{1994}).

\bibitem[{\citenamefont{Hill and Schauble}(2008)}]{Hill20081939}
\bibinfo{author}{\bibfnamefont{P.~S.} \bibnamefont{Hill}} \bibnamefont{and}
  \bibinfo{author}{\bibfnamefont{E.~A.} \bibnamefont{Schauble}},
  \bibinfo{journal}{Geochim. Cosmochim. Acta} \textbf{\bibinfo{volume}{72}},
  \bibinfo{pages}{1939 } (\bibinfo{year}{2008}).

\bibitem[{\citenamefont{Hrovat et~al.}(1997)\citenamefont{Hrovat, Hammons,
  Stevenson, and Borden}}]{Hrovat1997}
\bibinfo{author}{\bibfnamefont{D.~A.} \bibnamefont{Hrovat}},
  \bibinfo{author}{\bibfnamefont{J.~H.} \bibnamefont{Hammons}},
  \bibinfo{author}{\bibfnamefont{C.~D.} \bibnamefont{Stevenson}},
  \bibnamefont{and} \bibinfo{author}{\bibfnamefont{W.~T.}
  \bibnamefont{Borden}}, \bibinfo{journal}{J. Am. Chem. Soc.}
  \textbf{\bibinfo{volume}{119}}, \bibinfo{pages}{9523} (\bibinfo{year}{1997}).

\bibitem[{\citenamefont{Janak and Parkin}(2003)}]{Janak2003a}
\bibinfo{author}{\bibfnamefont{K.~E.} \bibnamefont{Janak}} \bibnamefont{and}
  \bibinfo{author}{\bibfnamefont{G.}~\bibnamefont{Parkin}},
  \bibinfo{journal}{Organometallics} \textbf{\bibinfo{volume}{22}},
  \bibinfo{pages}{4378} (\bibinfo{year}{2003}).

\bibitem[{\citenamefont{Kolmodin et~al.}(2002)\citenamefont{Kolmodin, Luzhkov,
  and Aqvist}}]{Kolmodin2002}
\bibinfo{author}{\bibfnamefont{K.}~\bibnamefont{Kolmodin}},
  \bibinfo{author}{\bibfnamefont{V.~B.} \bibnamefont{Luzhkov}},
  \bibnamefont{and} \bibinfo{author}{\bibfnamefont{J.}~\bibnamefont{Aqvist}},
  \bibinfo{journal}{J. Am. Chem. Soc.} \textbf{\bibinfo{volume}{124}},
  \bibinfo{pages}{10130} (\bibinfo{year}{2002}).

\bibitem[{\citenamefont{Munoz-Caro et~al.}(2003)\citenamefont{Munoz-Caro, Nino,
  Davalos, Quintanilla, and Abboud}}]{Munoz-Caro2003}
\bibinfo{author}{\bibfnamefont{C.}~\bibnamefont{Munoz-Caro}},
  \bibinfo{author}{\bibfnamefont{A.}~\bibnamefont{Nino}},
  \bibinfo{author}{\bibfnamefont{J.~Z.} \bibnamefont{Davalos}},
  \bibinfo{author}{\bibfnamefont{E.}~\bibnamefont{Quintanilla}},
  \bibnamefont{and} \bibinfo{author}{\bibfnamefont{J.~L.}
  \bibnamefont{Abboud}}, \bibinfo{journal}{J. Phys. Chem. A}
  \textbf{\bibinfo{volume}{107}}, \bibinfo{pages}{6160} (\bibinfo{year}{2003}).

\bibitem[{\citenamefont{Ruszczycky and Anderson}(2009)}]{Ruszczycky2009107}
\bibinfo{author}{\bibfnamefont{M.~W.} \bibnamefont{Ruszczycky}}
  \bibnamefont{and} \bibinfo{author}{\bibfnamefont{V.~E.}
  \bibnamefont{Anderson}}, \bibinfo{journal}{J. Mol. Struct. THEOCHEM}
  \textbf{\bibinfo{volume}{895}}, \bibinfo{pages}{107 } (\bibinfo{year}{2009}).

\bibitem[{\citenamefont{Saunders et~al.}(2007)\citenamefont{Saunders,
  Wolfsberg, Anet, and Kronja}}]{Saunders2007}
\bibinfo{author}{\bibfnamefont{M.}~\bibnamefont{Saunders}},
  \bibinfo{author}{\bibfnamefont{M.}~\bibnamefont{Wolfsberg}},
  \bibinfo{author}{\bibfnamefont{F.~A.~L.} \bibnamefont{Anet}},
  \bibnamefont{and} \bibinfo{author}{\bibfnamefont{O.}~\bibnamefont{Kronja}},
  \bibinfo{journal}{J. Am. Chem. Soc.} \textbf{\bibinfo{volume}{129}},
  \bibinfo{pages}{10276} (\bibinfo{year}{2007}).

\bibitem[{\citenamefont{Slaughter et~al.}(2000)\citenamefont{Slaughter,
  Wolczanski, Klinckman, and Cundari}}]{Slaughter2000}
\bibinfo{author}{\bibfnamefont{L.~M.} \bibnamefont{Slaughter}},
  \bibinfo{author}{\bibfnamefont{P.~T.} \bibnamefont{Wolczanski}},
  \bibinfo{author}{\bibfnamefont{T.~R.} \bibnamefont{Klinckman}},
  \bibnamefont{and} \bibinfo{author}{\bibfnamefont{T.~R.}
  \bibnamefont{Cundari}}, \bibinfo{journal}{J. Am. Chem. Soc.}
  \textbf{\bibinfo{volume}{122}}, \bibinfo{pages}{7953} (\bibinfo{year}{2000}).

\bibitem[{\citenamefont{Smirnov et~al.}(2009)\citenamefont{Smirnov, Lanci, and
  Roth}}]{Smirnov2009}
\bibinfo{author}{\bibfnamefont{V.~V.} \bibnamefont{Smirnov}},
  \bibinfo{author}{\bibfnamefont{M.~P.} \bibnamefont{Lanci}}, \bibnamefont{and}
  \bibinfo{author}{\bibfnamefont{J.~P.} \bibnamefont{Roth}},
  \bibinfo{journal}{J. Phys. Chem. A} \textbf{\bibinfo{volume}{113}},
  \bibinfo{pages}{1934} (\bibinfo{year}{2009}).

\bibitem[{\citenamefont{Zeller and Strassner}(2002)}]{Zeller2002}
\bibinfo{author}{\bibfnamefont{A.}~\bibnamefont{Zeller}} \bibnamefont{and}
  \bibinfo{author}{\bibfnamefont{T.}~\bibnamefont{Strassner}},
  \bibinfo{journal}{Organometallics} \textbf{\bibinfo{volume}{21}},
  \bibinfo{pages}{4950} (\bibinfo{year}{2002}).

\bibitem[{\citenamefont{Pollock and Ceperley}(1987)}]{Pollock1987}
\bibinfo{author}{\bibfnamefont{E.~L.} \bibnamefont{Pollock}} \bibnamefont{and}
  \bibinfo{author}{\bibfnamefont{D.~M.} \bibnamefont{Ceperley}},
  \bibinfo{journal}{Phys. Rev. B} \textbf{\bibinfo{volume}{36}},
  \bibinfo{pages}{8343} (\bibinfo{year}{1987}).

\bibitem[{\citenamefont{Boninsegni and Ceperley}(1995)}]{Boninsegni1995}
\bibinfo{author}{\bibfnamefont{M.}~\bibnamefont{Boninsegni}} \bibnamefont{and}
  \bibinfo{author}{\bibfnamefont{D.~M.} \bibnamefont{Ceperley}},
  \bibinfo{journal}{Phys. Rev. Lett.} \textbf{\bibinfo{volume}{74}},
  \bibinfo{pages}{2288} (\bibinfo{year}{1995}).

\bibitem[{\citenamefont{Torres et~al.}(2000)\citenamefont{Torres, Gelabert,
  Moreno, and Lluch}}]{TorresL._jp001327p}
\bibinfo{author}{\bibfnamefont{L.}~\bibnamefont{Torres}},
  \bibinfo{author}{\bibfnamefont{R.}~\bibnamefont{Gelabert}},
  \bibinfo{author}{\bibfnamefont{M.}~\bibnamefont{Moreno}}, \bibnamefont{and}
  \bibinfo{author}{\bibfnamefont{J.}~\bibnamefont{Lluch}}, \bibinfo{journal}{J.
  Phys. Chem. A} \textbf{\bibinfo{volume}{104}}, \bibinfo{pages}{7898}
  (\bibinfo{year}{2000}).

\bibitem[{\citenamefont{Ishimoto et~al.}(2008)\citenamefont{Ishimoto, Ishihara,
  Teramae, Baba, and Nagashima}}]{Ishimoto2008}
\bibinfo{author}{\bibfnamefont{T.}~\bibnamefont{Ishimoto}},
  \bibinfo{author}{\bibfnamefont{Y.}~\bibnamefont{Ishihara}},
  \bibinfo{author}{\bibfnamefont{H.}~\bibnamefont{Teramae}},
  \bibinfo{author}{\bibfnamefont{M.}~\bibnamefont{Baba}}, \bibnamefont{and}
  \bibinfo{author}{\bibfnamefont{U.}~\bibnamefont{Nagashima}},
  \bibinfo{journal}{J. Chem. Phys.} \textbf{\bibinfo{volume}{128}},
  \bibinfo{pages}{184309} (\bibinfo{year}{2008}).

\bibitem[{\citenamefont{Bardo and Wolfsberg}(1976)}]{Bardo1976}
\bibinfo{author}{\bibfnamefont{R.~D.} \bibnamefont{Bardo}} \bibnamefont{and}
  \bibinfo{author}{\bibfnamefont{M.}~\bibnamefont{Wolfsberg}},
  \bibinfo{journal}{J. Phys. Chem.} \textbf{\bibinfo{volume}{80}},
  \bibinfo{pages}{1068} (\bibinfo{year}{1976}).

\bibitem[{\citenamefont{Kleinman and Wolfsberg}(1973)}]{Kleinman1973}
\bibinfo{author}{\bibfnamefont{L.~I.} \bibnamefont{Kleinman}} \bibnamefont{and}
  \bibinfo{author}{\bibfnamefont{M.}~\bibnamefont{Wolfsberg}},
  \bibinfo{journal}{J. Chem. Phys.} \textbf{\bibinfo{volume}{59}},
  \bibinfo{pages}{2043} (\bibinfo{year}{1973}).

\bibitem[{\citenamefont{Bigeleisen}(1996)}]{Bigeleisen1996}
\bibinfo{author}{\bibfnamefont{J.}~\bibnamefont{Bigeleisen}},
  \bibinfo{journal}{J. Am. Chem. Soc.} \textbf{\bibinfo{volume}{118}},
  \bibinfo{pages}{3676} (\bibinfo{year}{1996}).

\bibitem[{\citenamefont{Knyazev et~al.}(1999)\citenamefont{Knyazev, Semin, and
  Bochkarev}}]{Knyazev19992579}
\bibinfo{author}{\bibfnamefont{D.~A.} \bibnamefont{Knyazev}},
  \bibinfo{author}{\bibfnamefont{G.~K.} \bibnamefont{Semin}}, \bibnamefont{and}
  \bibinfo{author}{\bibfnamefont{A.~V.} \bibnamefont{Bochkarev}},
  \bibinfo{journal}{Polyhedron} \textbf{\bibinfo{volume}{18}},
  \bibinfo{pages}{2579 } (\bibinfo{year}{1999}).

\bibitem[{\citenamefont{Cowan}(1981)}]{Cowan1981}
\bibinfo{author}{\bibfnamefont{R.~D.} \bibnamefont{Cowan}},
  \emph{\bibinfo{title}{The Theory of Atomic Structure and Spectra}}, Los
  Alamos Series in Basic and Applied Sciences (\bibinfo{publisher}{University
  of California Press}, \bibinfo{year}{1981}).

\bibitem[{\citenamefont{Doering and Zhao}(2006)}]{Doering2006}
\bibinfo{author}{\bibfnamefont{W.~V.} \bibnamefont{Doering}} \bibnamefont{and}
  \bibinfo{author}{\bibfnamefont{X.}~\bibnamefont{Zhao}}, \bibinfo{journal}{J.
  Am. Chem. Soc.} \textbf{\bibinfo{volume}{128}}, \bibinfo{pages}{9080}
  (\bibinfo{year}{2006}).

\bibitem[{\citenamefont{Doering and Keliher}(2007)}]{Doering2007}
\bibinfo{author}{\bibfnamefont{W.~V.} \bibnamefont{Doering}} \bibnamefont{and}
  \bibinfo{author}{\bibfnamefont{E.~J.} \bibnamefont{Keliher}},
  \bibinfo{journal}{J. Am. Chem. Soc.} \textbf{\bibinfo{volume}{129}},
  \bibinfo{pages}{2488} (\bibinfo{year}{2007}).

\bibitem[{\citenamefont{Chandler}(1987)}]{Chandler:1987}
\bibinfo{author}{\bibfnamefont{D.}~\bibnamefont{Chandler}},
  \emph{\bibinfo{title}{Introduction to modern statistical mechanics}}
  (\bibinfo{publisher}{Oxford University Press}, \bibinfo{address}{New York},
  \bibinfo{year}{1987}).

\bibitem[{\citenamefont{Feynman and Hibbs}(1965)}]{Feynman1965}
\bibinfo{author}{\bibfnamefont{R.}~\bibnamefont{Feynman}} \bibnamefont{and}
  \bibinfo{author}{\bibfnamefont{A.}~\bibnamefont{Hibbs}},
  \emph{\bibinfo{title}{Quantum Mechanics and Path Integrals}}, International
  Series in Pure and Applied Physics (\bibinfo{publisher}{McGraw-Hill},
  \bibinfo{year}{1965}).

\bibitem[{\citenamefont{Predescu et~al.}(2003)\citenamefont{Predescu, Sabo,
  Doll, and Freeman}}]{Predescu2003}
\bibinfo{author}{\bibfnamefont{C.}~\bibnamefont{Predescu}},
  \bibinfo{author}{\bibfnamefont{D.}~\bibnamefont{Sabo}},
  \bibinfo{author}{\bibfnamefont{J.~D.} \bibnamefont{Doll}}, \bibnamefont{and}
  \bibinfo{author}{\bibfnamefont{D.~L.} \bibnamefont{Freeman}},
  \bibinfo{journal}{J. Chem. Phys.} \textbf{\bibinfo{volume}{119}},
  \bibinfo{pages}{10475} (\bibinfo{year}{2003}).

\bibitem[{\citenamefont{Yamamoto and Miller}(2005)}]{yamamoto:044106}
\bibinfo{author}{\bibfnamefont{T.}~\bibnamefont{Yamamoto}} \bibnamefont{and}
  \bibinfo{author}{\bibfnamefont{W.~H.} \bibnamefont{Miller}},
  \bibinfo{journal}{J. Chem. Phys.} \textbf{\bibinfo{volume}{122}},
  \bibinfo{eid}{044106} (\bibinfo{year}{2005}).

\bibitem[{\citenamefont{Van\'{\i}\v{c}ek and Miller}(2007)}]{vanicek:114309}
\bibinfo{author}{\bibfnamefont{J.}~\bibnamefont{Van\'{\i}\v{c}ek}}
  \bibnamefont{and} \bibinfo{author}{\bibfnamefont{W.~H.}
  \bibnamefont{Miller}}, \bibinfo{journal}{J. Chem. Phys.}
  \textbf{\bibinfo{volume}{127}}, \bibinfo{eid}{114309} (\bibinfo{year}{2007}).

\bibitem[{\citenamefont{Wang and Zhao}(2009)}]{wang:114708}
\bibinfo{author}{\bibfnamefont{W.}~\bibnamefont{Wang}} \bibnamefont{and}
  \bibinfo{author}{\bibfnamefont{Y.}~\bibnamefont{Zhao}}, \bibinfo{journal}{J.
  Chem. Phys.} \textbf{\bibinfo{volume}{130}}, \bibinfo{eid}{114708}
  (\bibinfo{year}{2009}).

\bibitem[{\citenamefont{Van\'{\i}\v{c}ek
  et~al.}(2005)\citenamefont{Van\'{\i}\v{c}ek, Miller, Castillo, and
  Aoiz}}]{vanicek:054108}
\bibinfo{author}{\bibfnamefont{J.}~\bibnamefont{Van\'{\i}\v{c}ek}},
  \bibinfo{author}{\bibfnamefont{W.~H.} \bibnamefont{Miller}},
  \bibinfo{author}{\bibfnamefont{J.~F.} \bibnamefont{Castillo}},
  \bibnamefont{and} \bibinfo{author}{\bibfnamefont{F.~J.} \bibnamefont{Aoiz}},
  \bibinfo{journal}{J. Chem. Phys.} \textbf{\bibinfo{volume}{123}},
  \bibinfo{eid}{054108} (\bibinfo{year}{2005}).

\bibitem[{\citenamefont{Herman et~al.}(1982)\citenamefont{Herman, Bruskin, and
  Berne}}]{herman:5150}
\bibinfo{author}{\bibfnamefont{M.~F.} \bibnamefont{Herman}},
  \bibinfo{author}{\bibfnamefont{E.~J.} \bibnamefont{Bruskin}},
  \bibnamefont{and} \bibinfo{author}{\bibfnamefont{B.~J.} \bibnamefont{Berne}},
  \bibinfo{journal}{J. Chem. Phys.} \textbf{\bibinfo{volume}{76}},
  \bibinfo{pages}{5150} (\bibinfo{year}{1982}).

\bibitem[{\citenamefont{Van\'{\i}\v{c}ek and
  Miller}(2005)}]{Vanicek_Dubna_2005}
\bibinfo{author}{\bibfnamefont{J.}~\bibnamefont{Van\'{\i}\v{c}ek}}
  \bibnamefont{and} \bibinfo{author}{\bibfnamefont{W.~H.}
  \bibnamefont{Miller}}, in \emph{\bibinfo{booktitle}{Proceedings of the Eighth
  International Conference: Path Integrals from Quantum Information to
  Cosmology}}, edited by
  \bibinfo{editor}{\bibfnamefont{C.}~\bibnamefont{Burdik}},
  \bibinfo{editor}{\bibfnamefont{O.}~\bibnamefont{Navratil}}, \bibnamefont{and}
  \bibinfo{editor}{\bibfnamefont{S.}~\bibnamefont{Posta}}
  (\bibinfo{organization}{JINR Dubna}, \bibinfo{year}{2005}).

\bibitem[{\citenamefont{Predescu}(2004)}]{Predescu2004}
\bibinfo{author}{\bibfnamefont{C.}~\bibnamefont{Predescu}},
  \bibinfo{journal}{Phys. Rev. E} \textbf{\bibinfo{volume}{70}},
  \bibinfo{pages}{066705} (\bibinfo{year}{2004}).

\bibitem[{\citenamefont{Case et~al.}(2006)\citenamefont{Case, Darden,
  T.E.~Cheatham, Simmerling, J.~Wang, Luo, Merz, Pearlman, Crowley, Walker
  et~al.}}]{amber9}
\bibinfo{author}{\bibfnamefont{D.}~\bibnamefont{Case}},
  \bibinfo{author}{\bibfnamefont{T.}~\bibnamefont{Darden}},
  \bibinfo{author}{\bibfnamefont{I.}~\bibnamefont{T.E.~Cheatham}},
  \bibinfo{author}{\bibfnamefont{C.}~\bibnamefont{Simmerling}},
  \bibinfo{author}{\bibfnamefont{R.~D.} \bibnamefont{J.~Wang}},
  \bibinfo{author}{\bibfnamefont{R.}~\bibnamefont{Luo}},
  \bibinfo{author}{\bibfnamefont{K.}~\bibnamefont{Merz}},
  \bibinfo{author}{\bibfnamefont{D.}~\bibnamefont{Pearlman}},
  \bibinfo{author}{\bibfnamefont{M.}~\bibnamefont{Crowley}},
  \bibinfo{author}{\bibfnamefont{R.}~\bibnamefont{Walker}},
  \bibnamefont{et~al.}, \emph{\bibinfo{title}{Amber 9}} (\bibinfo{year}{2006}),
  \bibinfo{note}{\uppercase{U}niversity of California, San Francisco}.

\bibitem[{\citenamefont{Case et~al.}(2008)\citenamefont{Case, Darden,
  T.E.~Cheatham, Simmerling, Wang, Duke, Luo, Crowley, R.C.Walker, Zhang
  et~al.}}]{amber10}
\bibinfo{author}{\bibfnamefont{D.}~\bibnamefont{Case}},
  \bibinfo{author}{\bibfnamefont{T.}~\bibnamefont{Darden}},
  \bibinfo{author}{\bibfnamefont{I.}~\bibnamefont{T.E.~Cheatham}},
  \bibinfo{author}{\bibfnamefont{C.}~\bibnamefont{Simmerling}},
  \bibinfo{author}{\bibfnamefont{J.}~\bibnamefont{Wang}},
  \bibinfo{author}{\bibfnamefont{R.}~\bibnamefont{Duke}},
  \bibinfo{author}{\bibfnamefont{R.}~\bibnamefont{Luo}},
  \bibinfo{author}{\bibfnamefont{M.}~\bibnamefont{Crowley}},
  \bibinfo{author}{\bibnamefont{R.C.Walker}},
  \bibinfo{author}{\bibfnamefont{W.}~\bibnamefont{Zhang}},
  \bibnamefont{et~al.}, \emph{\bibinfo{title}{Amber 10}}
  (\bibinfo{year}{2008}), \bibinfo{note}{\uppercase{U}niversity of California,
  San Francisco}.

\bibitem[{\citenamefont{Chandler and Wolynes}(1981)}]{Chandler1981}
\bibinfo{author}{\bibfnamefont{D.}~\bibnamefont{Chandler}} \bibnamefont{and}
  \bibinfo{author}{\bibfnamefont{P.~G.} \bibnamefont{Wolynes}},
  \bibinfo{journal}{J. Chem. Phys.} \textbf{\bibinfo{volume}{74}},
  \bibinfo{pages}{4078} (\bibinfo{year}{1981}).

\bibitem[{\citenamefont{Roth and K\"{o}nig}(1966)}]{Roth1966}
\bibinfo{author}{\bibfnamefont{W.~R.} \bibnamefont{Roth}} \bibnamefont{and}
  \bibinfo{author}{\bibfnamefont{J.}~\bibnamefont{K\"{o}nig}},
  \bibinfo{journal}{J. Liebigs Ann. Chem.} \textbf{\bibinfo{volume}{699}},
  \bibinfo{pages}{24} (\bibinfo{year}{1966}).

\bibitem[{\citenamefont{Becke}(1997)}]{becke:8554}
\bibinfo{author}{\bibfnamefont{A.~D.} \bibnamefont{Becke}},
  \bibinfo{journal}{J. Chem. Phys.} \textbf{\bibinfo{volume}{107}},
  \bibinfo{pages}{8554} (\bibinfo{year}{1997}).

\bibitem[{\citenamefont{Schmider and Becke}(1998)}]{schmider:9624}
\bibinfo{author}{\bibfnamefont{H.~L.} \bibnamefont{Schmider}} \bibnamefont{and}
  \bibinfo{author}{\bibfnamefont{A.~D.} \bibnamefont{Becke}},
  \bibinfo{journal}{J. Chem. Phys.} \textbf{\bibinfo{volume}{108}},
  \bibinfo{pages}{9624} (\bibinfo{year}{1998}).

\bibitem[{\citenamefont{Dewar et~al.}(1985)\citenamefont{Dewar, Zoebisch,
  Healy, and Stewart}}]{Dewar1985}
\bibinfo{author}{\bibfnamefont{M.~J.~S.} \bibnamefont{Dewar}},
  \bibinfo{author}{\bibfnamefont{E.~G.} \bibnamefont{Zoebisch}},
  \bibinfo{author}{\bibfnamefont{E.~F.} \bibnamefont{Healy}}, \bibnamefont{and}
  \bibinfo{author}{\bibfnamefont{J.~J.~P.} \bibnamefont{Stewart}},
  \bibinfo{journal}{J. Am. Chem. Soc.} \textbf{\bibinfo{volume}{107}},
  \bibinfo{pages}{3902} (\bibinfo{year}{1985}).

\bibitem[{\citenamefont{Elstner et~al.}(1998)\citenamefont{Elstner, Porezag,
  Jungnickel, Elsner, Haugk, Frauenheim, Suhai, and Seifert}}]{Elstner1998}
\bibinfo{author}{\bibfnamefont{M.}~\bibnamefont{Elstner}},
  \bibinfo{author}{\bibfnamefont{D.}~\bibnamefont{Porezag}},
  \bibinfo{author}{\bibfnamefont{G.}~\bibnamefont{Jungnickel}},
  \bibinfo{author}{\bibfnamefont{J.}~\bibnamefont{Elsner}},
  \bibinfo{author}{\bibfnamefont{M.}~\bibnamefont{Haugk}},
  \bibinfo{author}{\bibfnamefont{T.}~\bibnamefont{Frauenheim}},
  \bibinfo{author}{\bibfnamefont{S.}~\bibnamefont{Suhai}}, \bibnamefont{and}
  \bibinfo{author}{\bibfnamefont{G.}~\bibnamefont{Seifert}},
  \bibinfo{journal}{Phys. Rev. B} \textbf{\bibinfo{volume}{58}},
  \bibinfo{pages}{7260} (\bibinfo{year}{1998}).

\bibitem[{\citenamefont{Merrick et~al.}(2007)\citenamefont{Merrick, Moran, and
  Radom}}]{Merrick2007}
\bibinfo{author}{\bibfnamefont{J.}~\bibnamefont{Merrick}},
  \bibinfo{author}{\bibfnamefont{D.}~\bibnamefont{Moran}}, \bibnamefont{and}
  \bibinfo{author}{\bibfnamefont{L.}~\bibnamefont{Radom}}, \bibinfo{journal}{J.
  Phys. Chem. A} \textbf{\bibinfo{volume}{111}}, \bibinfo{pages}{11683}
  (\bibinfo{year}{2007}).

\bibitem[{\citenamefont{Witek and Morokuma}(2004)}]{Witek2004}
\bibinfo{author}{\bibfnamefont{H.~A.} \bibnamefont{Witek}} \bibnamefont{and}
  \bibinfo{author}{\bibfnamefont{K.}~\bibnamefont{Morokuma}},
  \bibinfo{journal}{J. Comput. Chem.} \textbf{\bibinfo{volume}{25}},
  \bibinfo{pages}{1858} (\bibinfo{year}{2004}).

\bibitem[{\citenamefont{Malolepsza et~al.}(2005)\citenamefont{Malolepsza,
  Witek, and Morokuma}}]{Malolepsza2005}
\bibinfo{author}{\bibfnamefont{E.}~\bibnamefont{Malolepsza}},
  \bibinfo{author}{\bibfnamefont{H.~A.} \bibnamefont{Witek}}, \bibnamefont{and}
  \bibinfo{author}{\bibfnamefont{K.}~\bibnamefont{Morokuma}},
  \bibinfo{journal}{Chem. Phys. Lett.} \textbf{\bibinfo{volume}{412}},
  \bibinfo{pages}{237} (\bibinfo{year}{2005}).

\bibitem[{\citenamefont{Kr\"{u}ger et~al.}(2005)\citenamefont{Kr\"{u}ger,
  Elstner, Schiffels, and Frauenheim}}]{kruger:114110}
\bibinfo{author}{\bibfnamefont{T.}~\bibnamefont{Kr\"{u}ger}},
  \bibinfo{author}{\bibfnamefont{M.}~\bibnamefont{Elstner}},
  \bibinfo{author}{\bibfnamefont{P.}~\bibnamefont{Schiffels}},
  \bibnamefont{and}
  \bibinfo{author}{\bibfnamefont{T.}~\bibnamefont{Frauenheim}},
  \bibinfo{journal}{J. Chem. Phys.} \textbf{\bibinfo{volume}{122}},
  \bibinfo{eid}{114110} (\bibinfo{year}{2005}).

\bibitem[{\citenamefont{Flyvbjerg and Petersen}(1989)}]{flyvbjerg:461}
\bibinfo{author}{\bibfnamefont{H.}~\bibnamefont{Flyvbjerg}} \bibnamefont{and}
  \bibinfo{author}{\bibfnamefont{H.~G.} \bibnamefont{Petersen}},
  \bibinfo{journal}{J. Chem. Phys.} \textbf{\bibinfo{volume}{91}},
  \bibinfo{pages}{461} (\bibinfo{year}{1989}).

\bibitem[{\citenamefont{Berne and Thirumalai}(1986)}]{Berne1986}
\bibinfo{author}{\bibfnamefont{B.~J.} \bibnamefont{Berne}} \bibnamefont{and}
  \bibinfo{author}{\bibfnamefont{D.}~\bibnamefont{Thirumalai}},
  \bibinfo{journal}{Annu. Rev. Phys. Chem.} \textbf{\bibinfo{volume}{37}},
  \bibinfo{pages}{401} (\bibinfo{year}{1986}).

\bibitem[{\citenamefont{Martyna et~al.}(1992)\citenamefont{Martyna, Klein, and
  Tuckerman}}]{martyna:2635}
\bibinfo{author}{\bibfnamefont{G.~J.} \bibnamefont{Martyna}},
  \bibinfo{author}{\bibfnamefont{M.~L.} \bibnamefont{Klein}}, \bibnamefont{and}
  \bibinfo{author}{\bibfnamefont{M.}~\bibnamefont{Tuckerman}},
  \bibinfo{journal}{J. Chem. Phys.} \textbf{\bibinfo{volume}{97}},
  \bibinfo{pages}{2635} (\bibinfo{year}{1992}).

\bibitem[{\citenamefont{Wang et~al.}(2004)\citenamefont{Wang, Wolf, Caldwell,
  Kollman, and Case}}]{Wang2004}
\bibinfo{author}{\bibfnamefont{J.}~\bibnamefont{Wang}},
  \bibinfo{author}{\bibfnamefont{R.~M.} \bibnamefont{Wolf}},
  \bibinfo{author}{\bibfnamefont{J.~W.} \bibnamefont{Caldwell}},
  \bibinfo{author}{\bibfnamefont{P.~A.} \bibnamefont{Kollman}},
  \bibnamefont{and} \bibinfo{author}{\bibfnamefont{D.~A.} \bibnamefont{Case}},
  \bibinfo{journal}{J. Comput. Chem.} \textbf{\bibinfo{volume}{25}},
  \bibinfo{pages}{1157} (\bibinfo{year}{2004}).

\bibitem[{\citenamefont{Frisch et~al.}()\citenamefont{Frisch, Trucks, Schlegel,
  Scuseria, Robb, Cheeseman, Montgomery, Vreven, Kudin, Burant et~al.}}]{g03}
\bibinfo{author}{\bibfnamefont{M.~J.} \bibnamefont{Frisch}},
  \bibinfo{author}{\bibfnamefont{G.~W.} \bibnamefont{Trucks}},
  \bibinfo{author}{\bibfnamefont{H.~B.} \bibnamefont{Schlegel}},
  \bibinfo{author}{\bibfnamefont{G.~E.} \bibnamefont{Scuseria}},
  \bibinfo{author}{\bibfnamefont{M.~A.} \bibnamefont{Robb}},
  \bibinfo{author}{\bibfnamefont{J.~R.} \bibnamefont{Cheeseman}},
  \bibinfo{author}{\bibfnamefont{J.~A.} \bibnamefont{Montgomery},
  \bibfnamefont{Jr.}},
  \bibinfo{author}{\bibfnamefont{T.}~\bibnamefont{Vreven}},
  \bibinfo{author}{\bibfnamefont{K.~N.} \bibnamefont{Kudin}},
  \bibinfo{author}{\bibfnamefont{J.~C.} \bibnamefont{Burant}},
  \bibnamefont{et~al.}, \emph{\bibinfo{title}{Gaussian 03, \uppercase{R}evision
  \uppercase{E}.01}}, \bibinfo{note}{\uppercase{G}aussian, Inc., Wallingford,
  CT, 2004}.

\bibitem[{\citenamefont{Aradi et~al.}(2007)\citenamefont{Aradi, Hourahine, and
  Frauenheim}}]{AradiB._jp070186p}
\bibinfo{author}{\bibfnamefont{B.}~\bibnamefont{Aradi}},
  \bibinfo{author}{\bibfnamefont{B.}~\bibnamefont{Hourahine}},
  \bibnamefont{and}
  \bibinfo{author}{\bibfnamefont{T.}~\bibnamefont{Frauenheim}},
  \bibinfo{journal}{J. Phys. Chem. A} \textbf{\bibinfo{volume}{111}},
  \bibinfo{pages}{5678} (\bibinfo{year}{2007}).

\bibitem[{\citenamefont{Sunko et~al.}(1970)\citenamefont{Sunko, Humski,
  Malojcic, and Borcic}}]{SunkoDionisE._ja00725a026}
\bibinfo{author}{\bibfnamefont{D.~E.} \bibnamefont{Sunko}},
  \bibinfo{author}{\bibfnamefont{K.}~\bibnamefont{Humski}},
  \bibinfo{author}{\bibfnamefont{R.}~\bibnamefont{Malojcic}}, \bibnamefont{and}
  \bibinfo{author}{\bibfnamefont{S.}~\bibnamefont{Borcic}},
  \bibinfo{journal}{J. Am. Chem. Soc.} \textbf{\bibinfo{volume}{92}},
  \bibinfo{pages}{6534} (\bibinfo{year}{1970}).

\bibitem[{\citenamefont{Barborak et~al.}(1971)\citenamefont{Barborak, Chari,
  and Schleyer}}]{Barborak1971}
\bibinfo{author}{\bibfnamefont{J.~C.} \bibnamefont{Barborak}},
  \bibinfo{author}{\bibfnamefont{S.}~\bibnamefont{Chari}}, \bibnamefont{and}
  \bibinfo{author}{\bibfnamefont{P.~v.~R.} \bibnamefont{Schleyer}},
  \bibinfo{journal}{J. Am. Chem. Soc.} \textbf{\bibinfo{volume}{93}},
  \bibinfo{pages}{5275} (\bibinfo{year}{1971}).

\bibitem[{\citenamefont{Gajewski and Conrad}(1979)}]{Gajewski1979}
\bibinfo{author}{\bibfnamefont{J.~J.} \bibnamefont{Gajewski}} \bibnamefont{and}
  \bibinfo{author}{\bibfnamefont{N.~D.} \bibnamefont{Conrad}},
  \bibinfo{journal}{J. Am. Chem. Soc.} \textbf{\bibinfo{volume}{101}},
  \bibinfo{pages}{6693} (\bibinfo{year}{1979}).

\bibitem[{\citenamefont{Hascall et~al.}(1999)\citenamefont{Hascall, Rabinovich,
  Murphy, Beachy, Friesner, and Parkin}}]{Hascall1999}
\bibinfo{author}{\bibfnamefont{T.}~\bibnamefont{Hascall}},
  \bibinfo{author}{\bibfnamefont{D.}~\bibnamefont{Rabinovich}},
  \bibinfo{author}{\bibfnamefont{V.~J.} \bibnamefont{Murphy}},
  \bibinfo{author}{\bibfnamefont{M.~D.} \bibnamefont{Beachy}},
  \bibinfo{author}{\bibfnamefont{R.~A.} \bibnamefont{Friesner}},
  \bibnamefont{and} \bibinfo{author}{\bibfnamefont{G.}~\bibnamefont{Parkin}},
  \bibinfo{journal}{J. Am. Chem. Soc.} \textbf{\bibinfo{volume}{121}},
  \bibinfo{pages}{11402} (\bibinfo{year}{1999}).

\bibitem[{\citenamefont{Bender}(1995)}]{Bender1995}
\bibinfo{author}{\bibfnamefont{B.~R.} \bibnamefont{Bender}},
  \bibinfo{journal}{J. Am. Chem. Soc.} \textbf{\bibinfo{volume}{117}},
  \bibinfo{pages}{11239} (\bibinfo{year}{1995}).

\bibitem[{\citenamefont{Abu-Hasanayn et~al.}(1993)\citenamefont{Abu-Hasanayn,
  Krogh-Jespersen, and Goldman}}]{Abu-Hasanayn1993}
\bibinfo{author}{\bibfnamefont{F.}~\bibnamefont{Abu-Hasanayn}},
  \bibinfo{author}{\bibfnamefont{K.}~\bibnamefont{Krogh-Jespersen}},
  \bibnamefont{and} \bibinfo{author}{\bibfnamefont{A.~S.}
  \bibnamefont{Goldman}}, \bibinfo{journal}{J. Am. Chem. Soc.}
  \textbf{\bibinfo{volume}{115}}, \bibinfo{pages}{8019} (\bibinfo{year}{1993}).

\bibitem[{\citenamefont{Rabinovich and Parkin}(1993)}]{Rabinovich1993}
\bibinfo{author}{\bibfnamefont{D.}~\bibnamefont{Rabinovich}} \bibnamefont{and}
  \bibinfo{author}{\bibfnamefont{G.}~\bibnamefont{Parkin}},
  \bibinfo{journal}{J. Am. Chem. Soc.} \textbf{\bibinfo{volume}{115}},
  \bibinfo{pages}{353} (\bibinfo{year}{1993}).

\bibitem[{\citenamefont{Angus et~al.}(1935)\citenamefont{Angus, Bailey, Hale,
  Ingold, Leckie, Raisin, Thompson, and Wilson}}]{Angus1935}
\bibinfo{author}{\bibfnamefont{W.~R.} \bibnamefont{Angus}},
  \bibinfo{author}{\bibfnamefont{C.~R.} \bibnamefont{Bailey}},
  \bibinfo{author}{\bibfnamefont{J.~B.} \bibnamefont{Hale}},
  \bibinfo{author}{\bibfnamefont{C.~K.} \bibnamefont{Ingold}},
  \bibinfo{author}{\bibfnamefont{A.~H.} \bibnamefont{Leckie}},
  \bibinfo{author}{\bibfnamefont{C.~G.} \bibnamefont{Raisin}},
  \bibinfo{author}{\bibfnamefont{J.~W.} \bibnamefont{Thompson}},
  \bibnamefont{and} \bibinfo{author}{\bibfnamefont{C.~L.}
  \bibnamefont{Wilson}}, \bibinfo{journal}{J. Chem. Soc.}
  \textbf{\bibinfo{volume}{62}}, \bibinfo{pages}{971} (\bibinfo{year}{1935}).

\bibitem[{\citenamefont{Redlich}(1935)}]{Redlich1935}
\bibinfo{author}{\bibfnamefont{O.}~\bibnamefont{Redlich}}, \bibinfo{journal}{Z.
  Phys. Chem. Abt. B} \textbf{\bibinfo{volume}{28}}, \bibinfo{pages}{371}
  (\bibinfo{year}{1935}).

\bibitem[{\citenamefont{Stewart}(1989)}]{Stewart1989}
\bibinfo{author}{\bibfnamefont{J.~J.~P.} \bibnamefont{Stewart}},
  \bibinfo{journal}{J. Comput. Chem.} \textbf{\bibinfo{volume}{10}},
  \bibinfo{pages}{209} (\bibinfo{year}{1989}).

\bibitem[{\citenamefont{Rocha et~al.}(2006)\citenamefont{Rocha, Freire, Simas,
  and Stewart}}]{Rocha2006}
\bibinfo{author}{\bibfnamefont{G.~B.} \bibnamefont{Rocha}},
  \bibinfo{author}{\bibfnamefont{R.~O.} \bibnamefont{Freire}},
  \bibinfo{author}{\bibfnamefont{A.~M.} \bibnamefont{Simas}}, \bibnamefont{and}
  \bibinfo{author}{\bibfnamefont{J.~J.~P.} \bibnamefont{Stewart}},
  \bibinfo{journal}{J. Comput. Chem.} \textbf{\bibinfo{volume}{27}},
  \bibinfo{pages}{1101} (\bibinfo{year}{2006}).

\bibitem[{\citenamefont{McNamara et~al.}(2004)\citenamefont{McNamara, Muslim,
  Abdel-Aal, Wang, Mohr, Hillier, and Bryce}}]{McNamara2004}
\bibinfo{author}{\bibfnamefont{J.~P.} \bibnamefont{McNamara}},
  \bibinfo{author}{\bibfnamefont{A.-M.} \bibnamefont{Muslim}},
  \bibinfo{author}{\bibfnamefont{H.}~\bibnamefont{Abdel-Aal}},
  \bibinfo{author}{\bibfnamefont{H.}~\bibnamefont{Wang}},
  \bibinfo{author}{\bibfnamefont{M.}~\bibnamefont{Mohr}},
  \bibinfo{author}{\bibfnamefont{I.~H.} \bibnamefont{Hillier}},
  \bibnamefont{and} \bibinfo{author}{\bibfnamefont{R.~A.} \bibnamefont{Bryce}},
  \bibinfo{journal}{Chem. Phys. Lett.} \textbf{\bibinfo{volume}{394}},
  \bibinfo{pages}{429} (\bibinfo{year}{2004}).

\bibitem[{\citenamefont{Repasky et~al.}(2002)\citenamefont{Repasky,
  Chandrasekhar, and Jorgensen}}]{Repasky2002}
\bibinfo{author}{\bibfnamefont{M.~P.} \bibnamefont{Repasky}},
  \bibinfo{author}{\bibfnamefont{J.}~\bibnamefont{Chandrasekhar}},
  \bibnamefont{and} \bibinfo{author}{\bibfnamefont{W.~L.}
  \bibnamefont{Jorgensen}}, \bibinfo{journal}{J. Comput. Chem.}
  \textbf{\bibinfo{volume}{23}}, \bibinfo{pages}{1601} (\bibinfo{year}{2002}).

\bibitem[{\citenamefont{Stewart}(2007{\natexlab{a}})}]{Stewart2007}
\bibinfo{author}{\bibfnamefont{J.}~\bibnamefont{Stewart}}, \bibinfo{journal}{J.
  Mol. Model.} \textbf{\bibinfo{volume}{13}}, \bibinfo{pages}{1173}
  (\bibinfo{year}{2007}{\natexlab{a}}).

\bibitem[{\citenamefont{Stewart}(2007{\natexlab{b}})}]{mopac2007}
\bibinfo{author}{\bibfnamefont{J.~J.~P.} \bibnamefont{Stewart}},
  \emph{\bibinfo{title}{Mopac 2007}} (\bibinfo{year}{2007}{\natexlab{b}}),
  \bibinfo{note}{\uppercase{S}tewart Computational Chemistry, Colorado Springs,
  CO, USA}.

\bibitem[{\citenamefont{Liu et~al.}(1993)\citenamefont{Liu, Lynch, Truong, Lu,
  Truhlar, and Garret}}]{Liu1993}
\bibinfo{author}{\bibfnamefont{Y.~P.} \bibnamefont{Liu}},
  \bibinfo{author}{\bibfnamefont{G.~C.} \bibnamefont{Lynch}},
  \bibinfo{author}{\bibfnamefont{T.~N.} \bibnamefont{Truong}},
  \bibinfo{author}{\bibfnamefont{D.~H.} \bibnamefont{Lu}},
  \bibinfo{author}{\bibfnamefont{D.~G.} \bibnamefont{Truhlar}},
  \bibnamefont{and} \bibinfo{author}{\bibfnamefont{B.~C.}
  \bibnamefont{Garret}}, \bibinfo{journal}{J. Am. Chem. Soc.}
  \textbf{\bibinfo{volume}{115}}, \bibinfo{pages}{2408} (\bibinfo{year}{1993}).

\end{thebibliography}

\end{document}